\pdfoutput=1
\documentclass{sig-alternate}

\usepackage{xcolor,soul}
\usepackage{url}
\usepackage{cite}
\usepackage{xspace}
\usepackage{enumitem}
\usepackage{multirow}
\usepackage{amssymb}
\usepackage{comment}
\usepackage{rotating}
\usepackage{wrapfig}
\usepackage{balance}
\usepackage{mdwlist}

\newenvironment{bigtabular}[1]
{ \begin{tabular}{#1}}
{\end{tabular}}

\newcommand{\Wikipedia}{Wiki\-pedia\xspace}
\newcommand{\WPD}{HDP\xspace}
\newcommand{\WPDs}{HDPs\xspace}
\newcommand{\similarity}{sim_{\cos}}
\newcommand{\textSimilarity}{$\similarity$\xspace}
\newcommand{\noiseIndicator}[1]{_{#1\textit{noise}}}
\newcommand{\noiseProfile}[1]{e\noiseIndicator{#1}}
\newcommand{\noiseProfileIntersection}{\noiseProfile{{\bigcap}}}
\newcommand{\noiseProfileUnionAll}{\noiseProfile{{\bigcup}}}
\newcommand{\textNoiseProfile}{$\noiseProfile{}$\xspace}
\newcommand{\textIntersectionNoiseProfile}{$\noiseProfileIntersection$\xspace}
\newcommand{\textUnionAllNoiseProfile}{$\noiseProfileUnionAll$\xspace}
\newcommand{\mappingFunction}{m}
\newcommand{\textMappingFunction}{$\mappingFunction$\xspace}
\newcommand{\scoringFunction}{score}
\newcommand{\smoothedScoringFunction}{\scoringFunction\smoothingIndicator}
\newcommand{\similarityScoringFunction}{\similarity}
\newcommand{\textScoringFunction}{$\scoringFunction$\xspace}
\newcommand{\textSmoothedScoringFunction}{$\smoothedScoringFunction$\xspace}
\newcommand{\textSimilarityScoringFunction}{$\similarityScoringFunction$\xspace}
\newcommand{\weightingFunction}{w}
\newcommand{\smoothingIndicator}{'}
\newcommand{\smoothedWeightingFunction}{\weightingFunction\smoothingIndicator}
\newcommand{\tfidf}{\textit{tf}\scalebox{0.45}[1.0]{-}\textit{idf}}
\newcommand{\tfidfNorm}[1]{\left|\tfidf#1\right|}
\newcommand{\termFrequency}{\text{\textTermFrequency}}
\newcommand{\textWeightingFunction}{$\weightingFunction$\xspace}
\newcommand{\texttfidf}{$\tfidf$\xspace}
\newcommand{\textTFIDFNorm}{$\tfidfNorm{}$\xspace}
\newcommand{\textTermFrequency}{\textit{freq}\xspace}
\newcommand{\textKMeans}{\mbox{K-Means}\xspace}
\newcommand{\prob}{p}
\newcommand{\estimatedProb}{\hat{\prob}}
\newcommand{\bernoulliProb}{\estimatedProb_{\mathcal{B}}}
\newcommand{\multinomialProb}{\estimatedProb_{\mathcal{M}}}
\newcommand{\textNaiveBayes}{Na\"{i}ve Bayes\xspace}
\newcommand{\textBernoulliProb}{$\bernoulliProb$\xspace}
\newcommand{\textMultinomialProb}{$\multinomialProb$\xspace}
\newcommand{\maxLikelihoodIndicator}{_{_{ML}}}
\newcommand{\laplaceIndicator}{^{\mathcal{L}}}
\newcommand{\jelinekIndicator}{^{\mathcal{J}}}
\newcommand{\laplaceProb}{\estimatedProb\laplaceIndicator}
\newcommand{\maxLikelihoodProb}{\estimatedProb\maxLikelihoodIndicator}
\newcommand{\textLaplace}{Laplace\xspace}
\newcommand{\textJelinekMercer}{Jelinek-Mercer\xspace}
\newcommand{\bernoulliLaplaceProb}{\bernoulliProb\laplaceIndicator}
\newcommand{\multinomialJelinekProb}{\multinomialProb\jelinekIndicator}
\newcommand{\textBernoulliLaplaceProb}{$\bernoulliLaplaceProb$\xspace}
\newcommand{\textMultinomialJelinekProb}{$\multinomialJelinekProb$\xspace}
\newcommand{\fXMeasure}[1]{F_{#1}}
\newcommand{\fMeasure}{\fXMeasure{1}}
\newcommand{\microFMeasure}{micro(\fMeasure)}
\newcommand{\macroFMeasure}{macro(\fMeasure)}
\newcommand{\textFMeasure}{$\fMeasure$\xspace}
\newcommand{\textMicroFMeasure}{$\microFMeasure$\xspace}
\newcommand{\textMacroFMeasure}{$\macroFMeasure$\xspace}

\begin{document}
\title{Bootstrapped Grouping of Results \\to Ambiguous Person Name Queries}
\numberofauthors{4}
\author{
\alignauthor
Toni Gruetze\\
      \affaddr{Hasso Plattner Institute}\\
      \affaddr{Prof.-Dr.Helmert-Stra\ss{}e 2-3}\\
      \affaddr{14482 Potsdam, Germany}\\
      \email{toni.gruetze@hpi.uni-potsdam.de}
\alignauthor
Gjergji Kasneci\\
      \affaddr{Hasso Plattner Institute}\\
      \affaddr{Prof.-Dr.Helmert-Stra\ss{}e 2-3}\\
      \affaddr{Potsdam, Germany}\\
      \email{gjergji.kasneci@hpi.uni-potsdam.de}
\alignauthor
Zhe Zuo\\
      \affaddr{Hasso Plattner Institute}\\
      \affaddr{Prof.-Dr.Helmert-Stra\ss{}e 2-3}\\
      \affaddr{Potsdam, Germany}\\
      \email{zhe.zuo@hpi.uni-potsdam.de}\\
\and
\alignauthor
Felix Naumann\\
      \affaddr{Hasso Plattner Institute}\\
      \affaddr{Prof.-Dr.Helmert-Stra\ss{}e 2-3}\\
      \affaddr{Potsdam, Germany}\\
      \email{felix.naumann@hpi.uni-potsdam.de}
}
\additionalauthors{}
\makeatletter
  \def\@copyrightspace{\relax}
  \makeatother

\maketitle


\begin{abstract}
\begin{bf}
Some of the main ranking features of today's search engines reflect result popularity and are based on ranking models, such as PageRank, implicit feedback aggregation, and more. While such features yield satisfactory results for a wide range of queries, they aggravate the problem of search for ambiguous entities: Searching for a person yields satisfactory results only if the person we are looking for is represented by a high-ranked Web page and all required information are contained in this page. Otherwise, the user has to either reformulate/refine the query or manually inspect low-ranked results to find the person in question. A possible approach to solve this problem is to cluster the results, so that each cluster represents one of the persons occurring in the answer set. However clustering search results has proven to be a difficult endeavor by itself, where the clusters are typically of moderate quality.

A wealth of useful information about persons occurs in Web~2.0 platforms, such as LinkedIn, \Wikipedia, Facebook, etc. Being human-generated, the information on these platforms is clean, focused, and already disambiguated. We show that when searching for ambiguous person names the information from such platforms can be bootstrapped to group the results according to the individuals occurring in them. We have evaluated our methods on a hand-labeled dataset of around 5,000 Web pages retrieved from Google queries on 50 ambiguous person names.
\end{bf}
\end{abstract}


\section{Introduction}
With ever more information being placed on the Web, established retrieval techniques are undergoing a stress test. Although search engines have matured by integrating different relevance criteria, e.g., query-based and social relevance, result freshness, user interests, etc., they still lack the ability to effectively respond to ambiguous queries for specific entities, such as people, products, locations, etc. The common way by which modern search engines approach the ambiguity problem is by diversifying search results and hoping that at least one of the top-10 results satisfies the user's information need. However, the employed ranking strategies mostly rely on authority- and popularity-based measures, e.g., PageRank scores~\cite{Page1999}, models for aggregating implicit relevance feedback (e.g., in terms of user clicks), etc. As a consequence, while the diversification approach works well for popular searches, for which there exist authoritative Web pages and plenty of user feedback, there is a long tail of results to ambiguous queries, which does not fulfill the mentioned criteria. The severeness of this problem becomes especially obvious in search tasks involving ambiguous person names. In such cases, the returned results are only satisfactory if the person in question is represented by a high-ranked Web page with the required information. Otherwise, the user has to either refine the query (through additional terms) or manually inspect low-ranked results.

For example, suppose that you have recently attended a conference talk by computer scientist Michael Jordan and you are interested in more background information about the speaker and his research. Searching for the name ``Michael Jordan'' on Google yields top-10 results entirely about the former basketball player.In fact, the first page about the well-known researcher from the U.C.~Berkeley, is ranked $18^{\text{th}}$ in the result list\footnote{The second hit about the researcher is ranked $47^{\text{th}}$.}. For the average user, who aims at a top-10 hit for his search, this is unacceptable.

For such queries, it would be useful to present the user with clusters of results where each cluster represents one of the individuals occurring in the answer set. Typical approaches to this problem retrieve salient text snippets for the ambiguous query and cluster results based on textual similarity measures, using predefined or learned thresholds. Other features, such as links or word senses (concepts), can also be taken into account. Obviously, such techniques have to handle a lot of noise and it is questionable whether they can handle highly ambiguous person-related queries (e.g., ``Mi\-chael Jordan''), with different persons of the same name and the same category, say ``computer scientists'' (DBLP alone lists 3 Mi\-chael Jordan). Under such noisy conditions typical clustering techniques, such as \textKMeans or Hierarchical Agglomerative Clustering (HAC)~\cite{Manning2008}, are shown to perform rather moderately~\cite{Zamir1998}. We have evaluated the performance of well-known clustering methods, such as HAC and \textKMeans on our dataset (see Section~\ref{sec:evaluation:results:clustering}), and found that although they perform well in terms of purity, but the resulting clusters yield low normalized mutual information (NMI) scores. Our approaches outperformed both methods by approximately $10\%$ in terms of purity, respectively up to $95\%$ in terms of NMI.

A study on the distribution of query types on AlltheWeb and AltaVista shows that, already in 2003, person-related queries made up a non-negligible portion of the Web search queries~\cite{Spink2004}. In today's Web ever more information centers around people, be it in social networks, encyclopedic sources, or professional homepages. LinkedIn, Facebook, Wikipedia, Twitter, and Xing are only part of a wide and continuously growing range of Web~2.0 platforms centering around people and other entities. Therefore, we believe that person-related queries will continue to gain importance. At the same time, being human-generated, the information on such platforms is focused and already disambiguated~\cite{Berendsen2012}.

Our approach to the ambiguous person search problem can bootstrap the information of a knowledge base to identify groups of results that represent unique individuals. In this way the original clustering problem is cast into a classification problem, where the classes are given by the different same-name individuals occurring in the knowledge base. We are aware of the existing coverage problem: there remain many people who may not appear in the knowledge base. However, as shown in our evaluation (i.e., Section~4.2) the classification into knowledge base entities leads to a quality enhancement for pages regarding these particular entities. Furthermore, the remaining pages might be clustered traditionally or further linked to entities of other knowledge bases in a separate step, but this is not in the scope of this work.

Note that for any practical relevance, approaches to ambiguous person search on the Web need to be effective (i.e., the returned groups need to have high quality) and efficient (i.e., the grouping of results should happen in an online fashion, avoiding long waiting times for the user). Other non-obvious requirements are that the system should handle noise and uncertainty in the grouping process. In this paper, we analyze the result quality of different efficient Information Retrieval and Machine Learning strategies to solve the above problem and show that information bootstrapped from Web~2.0 sources can considerably improve the result of the disambiguation process.

In summary, our contributions are the following:
\begin{enumerate}
\item We propose a framework for transforming the task of clustering results to ambiguous person name queries into a classification task that builds on bootstrapping knowledge base entities.
\item We propose and investigate different strategies for mitigating bias and noise during the disambiguation process. While bias inevitably arises from a document ranking (which, in case of PageRank, yields top results of highly linked Web pages about few more ``Web-popular'' individuals and many low-rank ones about many others), noise may arise from single result pages containing information about different people or about people who are not represented in the knowledge base (open world assumption).
\item We investigate the quality of different efficient Informa\-tion Retrieval and Machine Learning algorithms with respect to the result disambiguation problem.
\item We demonstrate the viability of our approach in experiments on a hand-labeled dataset of around 5,000 Web pages\footnote{Available at \url{http://hpi-web.de/naumann/projekte/repeatability/datasets/wpsd.html}} retrieved from Google queries on 50 ambiguous person names appearing on ``Wikipedia's Human Disambiguation Pages''.
\end{enumerate}

The remainder of this document is organized as follows: Section~\ref{sec:related_work} discusses related research. Section~\ref{sec:algorithm} introduces our Web page classification approaches and Section~\ref{sec:evaluation} discusses their experimental evaluation. Finally, we discuss future work and conclude in Section~\ref{sec:conclusion}.


\section{Background and Related Work}
\label{sec:related_work}
Entity disambiguation is a broad topic and spans several well-studied research fields in computer science~\cite{Fellegi1969}. Due to the large amount of scientific publications in this area, we can discuss only the most relevant fraction of the existing related work with no claim for completeness. In the following we discuss three different research streams: Entity Resolution, Entity Linking, and Text Clustering.

The field of \textbf{Entity Resolution} (ER) (also referred to as record linkage, deduplication, or data matching by the database community) is already summarized in several survey works \cite{Christen2012a,Christen2012b,Elmagarmid2007,Kopcke2010,Naumann2010}. In this realm, the focus is on comparing sets of entities and identifying convincing/correct mappings between them. However, ER methods typically assume structured entity data, such as database entries with a defined set of attributes (commonly with a value range), which we cannot assume.

\textbf{Entity Linking} (EL) is the task of linking mentions of named entities in Web text with their referent entities in a knowledge base. These knowledge bases might be extracted from various sources, such as \Wikipedia, DBLP, IMDb, etc. For instance, Bunescu and Pasca transform the named entity disambiguation problem to a ranking problem of \Wikipedia entities~\cite{Bunescu2006}. They derive an entity dictionary from \Wikipedia and rank these entities for a given mention according to a scoring function that is based the cosine similarity between the textual context of the mention and \Wikipedia's text and categories of the candidate entity. Cucerzan~\cite{Cucerzan2007} extends the feature set of a candidate \Wikipedia entity by information from other articles linking to the candidate's article, but instead of using the whole article text only some key phrases and immediate categories are included. Fader et al.~\cite{Fader2009} extend the previous works by factoring a prominence prior into the disambiguation decision.

Hassell et al.\ disambiguate the names of academic researchers included in a collection of DBWorld posts by analyzing relationship information between research articles, computer scientists, journals, and proceedings that are extracted from DBLP~\cite{Hassell2006}. All these techniques address a problem similar to ours. However, our problem has to be solved by more efficient methods, because they have to be applicable to online scenarios and process complete Web pages, which are results to a Web search query.

Another related research field is \textbf{Text Clustering}. The focus is on grouping texts about ambiguous named entities, such that every group uniquely represents an individual entity. Early works in this realm use clustering for cross-document co-reference resolution, i.e., to find referents across multiple documents~\cite{Bagga1998}. Subsequent works cluster Web appearances of ambiguous named entities and try to identify the underlying real world entity (e.g.,~\cite{Mann2003}). One of the most challenging tasks here is the disambiguation of search results to ambiguous person names~\cite{Artiles2010,Artiles2007,Artiles2009}, also referred to as personal name resolution~\cite{Balog08}.

For text clustering, a wide variety of feature selection strategies can be observed: A common representation of Web documents is given by the vector space model of the document terms \cite{Bagga1998,Bekkerman2005,Berendsen2012,Long2010,Smirnova2010,Vu2007,Wan2005}. The features are typically weighted by their \texttfidf weights. Further, more advanced features can be considered, such as named entities \cite{Ikeda2009,Kalashnikov2007,Long2010,Wan2005}, noun phrases \cite{Ikeda2009,Wan2005}, intrinsic hyper-link relationships among Web pages \cite{Bekkerman2005,Ikeda2009,Kalashnikov2007,Smirnova2010}, email address references \cite{Kalashnikov2007,Wan2005}, and detailed personal or biographical features, such as occupation and title, phone number, or birth year \cite{Mann2003,Wan2005}. Such features typically lead to a high precision but a low recall of results, because ``they are not observed frequently, but work as strong evidence''\cite{Ikeda2009}. Balog et al. compare the performance of a simple bag-of-word based clustering approaches for the personal name resolution task and showed comparable results to state-of-the-art approaches that base on more sophisticated features~\cite{Balog08}. Furthermore, Sekine and Artiles state that the extraction of some of these ``advanced features'' (such as birth date, spouse name, occupation, \ldots) is especially hard and that existing solutions generally provide low \textFMeasure measure scores~\cite{Sekine2009}. This poor performance occurs, because Web documents (in contrast to scientific texts, or encyclopedic articles) can be very noisy.

Once the features are selected, the approaches employ various clustering methods to achieve the final grouping. A commonly used clustering method is Hierarchical Agglomerative Clustering (HAC) \cite{Berendsen2012,Long2010,Mann2003,Smirnova2010,Vu2007,Wan2005}. Based on this technique, Ikeda et al.\ introduce a two-stage clustering~\cite{Ikeda2009}.
In the first stage, HAC creates clusters that are used to extract refined features for the second stage clustering. Bekkerman and McCallum provide an agglomerative/conglomerative double clustering method~\cite{Bekkerman2005}. It can be shown that this technique is related to the information bottleneck method, which is known to perform well for text clustering tasks~\cite{Slonim2000}. As one can see, hierarchical algorithms are commonly used in this research field. It is a widespread belief hat hierarchical algorithms have a higher clustering quality than partitional methods (e.g., \textKMeans), which typically provide a better run-time behavior in high dimensional feature spaces. However, Zhao and Karypis showed that partitional methods can outperform hierarchical methods in terms of clustering quality~\cite{Zhao2002}. Pedersen et al.\ apply the ``Repeated Bisections'' of \cite{Zhao2002}, a hybrid approach that combines \textKMeans and hierarchical divisive clustering~\cite{Pedersen2005}. They cluster documents based on second order context vectors. These vectors are based on a bigram matrix that is reduced through Singular Value Decomposition. The authors of~\cite{Berendsen2012} introduce a three-step clustering algorithm that makes use of social network profiles from various networks and is in this respect in the spirit of our approach. In the first step the profiles are clustered and in the second step the result documents are clustered. Finally, the profile clusters are merged with the document clusters. In contrast to this approach, we avoid the noise of clustering by bootstrapping social profiles as clean seeds against which Web pages have to be matched.

In the evaluation section, we compare our methods with traditional Text Clustering techniques. The results show that the bootstrapping of the knowledge bases leads to results that are superior to those returned by unsupervised techniques.


\section{Web page classification}
\label{sec:algorithm}
As stated earlier, our goal is to transform the clustering of search results to ambiguous person name queries into a classification task, by bootstrapping knowledge base entities about people with the same name. To this end we define the disambiguation task as follows:

For an ambiguous person name $x$, let $D_x = \{d_1, \ldots, d_n\}$ denote the set of retrieved documents to the query $x$, e.g., Google search results to the query $x$. Furthermore, for an entity source $S$  let $E_x (S) = \{e_1, \ldots, e_m\}$ be the set of entity profiles in $S$ that are referred to by the same name $x$, e.g., those entities could be retrieved from \Wikipedia disambiguation pages or from a name search API in case of other entity sources, such as LinkedIn or Facebook. The task we address is the construction of a surjective mapping $\mappingFunction_x : D_x \rightarrow E_x (S)\cup \{\noiseProfile{}\}$ such that a document $d\in D_x$ is mapped to $e\in E_x(S)$ if and only if $d$ is about $e$, and to a ``noise entity'' $\noiseProfile{}$ if $d$ does not describe any of the entities in $E_x(S)$.

In the following, for the sake of a simpler notation, we omit the index $x$ and the symbol $S$ of the entity source and assume them to be implicitly given by the context. Also, for compactness, we use the term ${E}'$ to denote the extended entity profile set (${E}' = E\cup \{\noiseProfile{}\}$). The definition and construction of these noise profiles is explained in Section~\ref{sec:algorithm:noise_profiles}.

The above problem could be modeled as a graph partitioning problem, where entities and result pages would be connected by weighted edges representing their similarities.
\begin{wrapfigure}{r}{.45\linewidth}
\centering
\includegraphics[trim=75 50 225 40, clip=true, width=65pt]{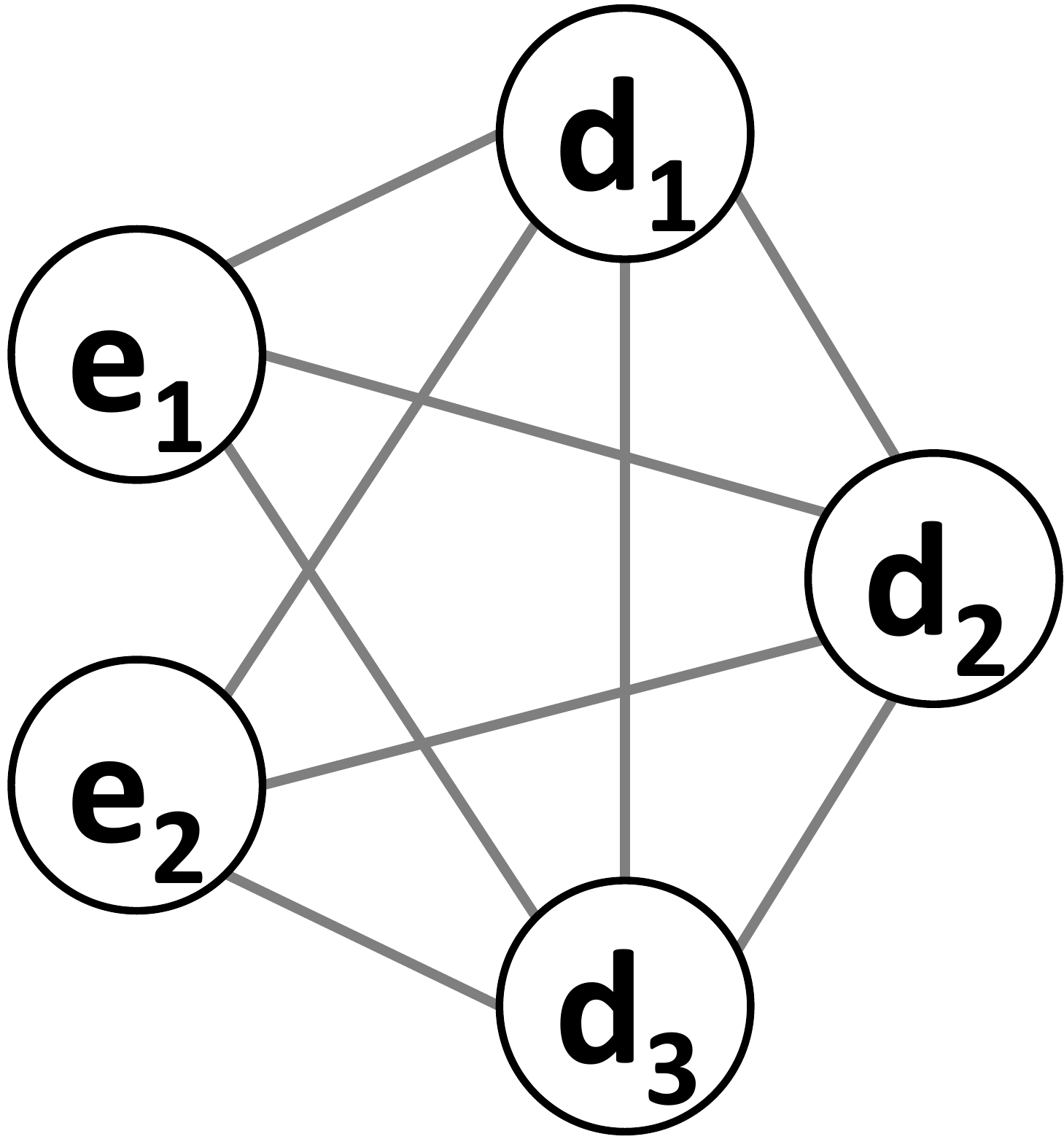}
\vspace{-10pt}
\caption{Example graph for the partitioning problem.}
\label{fig:algorithm:matching_problem}
\end{wrapfigure}
Figure~\ref{fig:algorithm:matching_problem} shows an example graph, where edges indicate differently weighted similarity relationships. Note that there is no edge between $e_1$ and $e_2$ ($e_1,e_2 \in {E}'$), because we apply a disjointness constraint between all entities from $E'$. The objective then would be to partition the graph into $|E'|$ components while minimizing the total weight of the edges between separate components. This problem is known to be NP-hard~\cite{Watrigant2012}; hence we follow a relaxed version of this model, where we assign each document to the entity partition for which it exhibits the largest similarity or probability score. To this end, we propose a suite of vector space and probabilistic models, which can be analogously applied to address the above problem.

Note that Web page clustering enables the user to browse the results in a more efficient way~\cite{Zamir1999}. However, it remains to mention that in general, a clustering of search results bears significant risks, since, if a result document was assigned to the wrong cluster, it would be difficult for the user to find it. To mitigate this risk, we propose to show the clusters in addition to the ranked search results as in~\cite{Kalashnikov2007}. If the user explores a specific entity cluster, corresponding to an entity $e_i \in E'$, we first present all result documents $ d_j \in D$ for which $m(d_j) = e_i$ holds and list subsequently all other result documents (i.e., sorted by a similarity score). Thus, the user has the chance to examine all related Web pages.

\subsection{Vector Space Model}
\label{sec:algorithm:vector_space_model}
Let ${C}' = D \cup {E}'$ denote the corpus of documents and profiles about entities that are referred to by the same name. Let $F_c$ denote the features of a document $c\in C'$. The feature space
\[
F = \bigcup\limits_{c \in {C}'} F_c
\]
contains all features from $C'$.

Being relatively succinct, an entity profile might miss many features that could also be salient for the corresponding entity. These missing features, however, could be found in other Web documents about the same entity. Hence, in our model the \textbf{similarity score} between an entity profile $e \in {E}'$ and a result document $d\in D$ is captured not only by the features that they actually have in common but also by the features that they might have in common if the profile was fully extended. More specifically, we define
\[
	\scoringFunction(d, e) = \sum\limits_{f \in F} \weightingFunction(f, e) \cdot \weightingFunction(f, d)
\]
where $\weightingFunction$ should reflect the importance of feature $f$ for an element $c\in {C}'$ and can be modeled in different ways, e.g., as a metric distance, an information-theoretic similarity measure, a probabilistic measure, etc. We have analyzed the result quality for many different options and present the ones that turned out to be most promising in our experiments. Thereby, we used the well known cosine similarity (\textSimilarityScoringFunction) as a \textbf{baseline scoring function}.

Note that in general, the features could be of various types; they could be source dependent (e.g., semi-structured or multimodal sources suggest other features than unstructured sources), and they could also be more complex in nature by involving inter-document-links, $n$-grams ($n>1$), named entities, factual phrases, compound nouns, etc. However, for most online scenarios, the challenge is to exploit simple features and yet return high-quality clusters of results. The evaluation of our bootstrapping approach (see Section~\ref{sec:evaluation:results}) shows that it is possible to rely on word-unigrams and still provide high-quality results. However, especially semi-structured information sources might additionally provide richer features like factual phrases regarding particular attributes (i.e., research topics and working places in LinkedIn profiles) and thus enhance the result quality~\cite{Sekine2009}.

Finally, the mapping function $m: D \rightarrow {E}'$ assigns a document $d\in D$ to the entity profile $e\in {E}'$ that maximizes the similarity score $\scoringFunction(d,e)$ ($\similarityScoringFunction(d,e)$, respectively).

\subsubsection*{Measures for the Weight Function \textWeightingFunction}
As mentioned above, the similarity between an entity profile and a result document is measured over the features they have or might not have in common. These kinds of explicit and implicit similarities are incorporated into our model by different choices of the weight function $\weightingFunction$.

A popular measure for the importance of a feature for a given document is the $\texttfidf$ measure. For the explicit similarity between entity profiles and result documents we define the weight measure for a feature $f\in F$ and an element $c\in {C}'$ simply as
\[
	\weightingFunction(f,c)=\tfidf(f,c)
\]
where implementation-wise we have chosen
\[
	\tfidf(f,c)=\frac{\termFrequency(f, c)}{\max\limits_{f' \in F_{c}}\termFrequency(f', c)} \cdot \log\left(\frac{|F_{c}|}{|\{c' \in C: f \in F_{c'}\}|}\right).
\]

For measuring the implicit similarity between entity profiles and result documents, we define the smoothed weight function for the importance of a feature $f\in F$ for an entity profile $e\in {E}'$ as
\[
	\smoothedWeightingFunction(f,e)=\tfidfNorm{(f,e)}+\sum_{d\in D} \similarity(e,d)\cdot\tfidfNorm{(f,d)}
\]
where \textSimilarity stands for the cosine similarity between the weighted feature vector of the entity profile $e$ and that of the document $d$, and \textTFIDFNorm represents the L1-normalized \texttfidf score. The final \textbf{smoothed similarity score} is given by:
\[
\smoothedScoringFunction(d, e) = \sum\limits_{f \in F} \smoothedWeightingFunction(f, e) \cdot \weightingFunction(f, d)
\]
The intuition behind this measure is that the normalized \texttfidf-based feature vector representing the entity profile is first `pulled' towards similar document vectors and then its (inner-product-based) similarity to the document vector is computed. This method is similar in spirit to the Rocchio algorithm for modifying the original query vector when relevance feedback is available~\cite{Manning2008}. Note that the L1-normalization allows us to move the entity profile vector fraction-wise towards similar document vectors. This accounts for a careful (and rather conservative) modification of the original entity profile vector. The outcome of the final inner-product is proportional to the cosine similarity between the modified vector and the document vector.

\subsection{Probabilistic Models}
\label{sec:algorithm:probabilistic_model}
We now discuss the application of probabilistic models to our classification problem. The mapping $m: D \rightarrow {E}'$ maps a document $d\in D$ to the entity profile $e\in {E}'$ that maximizes the joint probability $\prob(d,e)$.

By applying the chain rule and assuming conditional feature independence for a given entity-profile, we can derive a \textbf{Bernoulli \textNaiveBayes} model (\textBernoulliProb)
\[
	\bernoulliProb(e, d) = \prob(e) \cdot \prod\limits_{f \in F_d}\prob(f | e)
\]
where $\prob(e)$ is a prior describing the prominence of the entity represented by the profile $e$, and $\prob(d | e)$ captures the plausibility of the document $d$ being generated from the entity profile $e$. As an alternative, we consider a \textbf{Multinomial \textNaiveBayes} model (\textMultinomialProb), which takes feature occurrence frequencies into account:
\[
\multinomialProb(e, d) = |d|!\cdot\prob(e)\cdot \prod\limits_{f \in F_d} \frac{\prob(f | e)^{\termFrequency(f, d)}}{\termFrequency(f, d)!}
\]
where $\termFrequency(f, d)$ represents the absolute frequency of the feature $f$ in the document $d$, so that $\sum_{f\in F_d} \termFrequency(f, d) = |d|$.

\subsubsection*{Parameter Estimation}
Due to the confined nature of the feature set of a given entity profile, simple maximum likelihood estimation of the conditionals $\prob(f | e)$ would not be appropriate and lead to underestimations. Furthermore, the model would be prone to numerical effects, especially for cases where $f\notin F_e$, i.e., the feature $f$ does not occur in the entity profile $e$. A possible solution to this problem is the extension of the feature set of $e$ with features from the documents similar to the actual entity~\cite{Efron2012,Liu2004,Tao2006}. We implemented models that extended the feature sets with features from the top-$k$ documents that are ``closest'' to $e$ (e.g., in terms of cosine similarity, Jaccard distance, etc.) and experimented with different values for $k$. However, the results were fairly disappointing (both in terms of precision and recall measures). Seemingly, the model could not handle the noise introduced by the extension of the feature set. Much better results were achieved by the following simple smoothing techniques

\textbf{\textLaplace}-smoothing (also referred to as additive smoothing) adds a smoothing factor $\alpha$ to the actual relative frequency of each feature. Thus, the prior is estimated by:
\[
\laplaceProb_{\weightingFunction}(e) = \frac{\sum\limits_{f \in F_e}\weightingFunction(f, e)+\alpha}{\sum\limits_{e_j \in E}\sum\limits_{f_i \in F_{e_j}}\weightingFunction(f_i, e_j)+\alpha}
\]
and the likelihood is defined as:
\[
\laplaceProb_{\weightingFunction}(f | e) = \frac{\weightingFunction(f, e)+\alpha}{\sum\limits_{f_i \in F_e}\weightingFunction(f_i, e)+\alpha}
\]
In our experiments, a smoothing parameter $\alpha = 0.01$ empirically showed best results.

Another popular smoothing method, the \textbf{\textJelinekMercer}-smooth\-ing, uses a background model (based on corpus frequencies) to estimate the likelihood of non-occurring features. It is defined as:
\[
\estimatedProb_{\lambda}(f | e) = (1-\lambda)\maxLikelihoodProb(f | F_e) + \lambda\prob(f | F)
\]
It can be shown that by setting $\lambda$ to 0.5, one can derive a \texttfidf-style smoothing, which we used in our implementation. In our experiments we found that the Bernoulli \textNaiveBayes model worked best with Laplace smoothing (\textBernoulliLaplaceProb), while the Multinomial model worked best with the Jelinek-Mercer smoothing (\textMultinomialJelinekProb). We report the corresponding results in  Section~\ref{sec:evaluation:results}.

\subsection{Modeling the Noise Entity Profile}
\label{sec:algorithm:noise_profiles}
In the definition of our mapping \textMappingFunction we introduced an artificial entity profile $\noiseProfile{}$, to which documents should be mapped if they do not match any of the entity profiles in $E$. This addition accounts for the fact that the set of unique entities having the same name is limited by the underlying bootstrapping source. Hence, result documents that do not correspond to any of the entity profiles from $E$ are assigned to the artificial profile $\noiseProfile{}$. There are different ways to model such a noise profile; in general, however, it should contain rather uninformative features, e.g., features with low expected information gain or features with high $df$ values. We tested various approaches, but for the sake of brevity, we present the two best performing and most robust ones.

As a first approach we consider the \textbf{union-noise entity} profile, denoted by $\noiseProfileUnionAll$. The profile is generated in a straightforward manner by equally weighting all entity features.
\[
F_{\noiseProfileUnionAll}=\{f | f\in F_e, e\in E\}
\]
This method aims at maximizing the feature noise in the artificial profile.

In addition to $\noiseProfileUnionAll$ we introduce the \textbf{intersection-noise entity} profile, denoted by $\noiseProfileIntersection$. It contains all features (equally weighted) occurring in the intersection of any entity profile $e$ with any document $c$ from the corpus:
\[
F_{\noiseProfileIntersection}=\{ f | \forall e\in E, \forall c \in C, e\neq c: f \in F_e \cap F_c \},
\]
where $C = C'\setminus\{\noiseProfile{}\}$. Note that the above definition of the noise entity is biased towards features with high $\textit{df}$ values (i.e., non-specific features), but may still contain slightly informative features, thus mitigating a rigid discrimination between the noise entity and the other entity profiles.

We have evaluated the effect of both approaches on the mapping quality and show the results in the next section.


\section{Evaluation}
\label{sec:evaluation}
The problem of clustering search results to ambiguous person name queries has been studied in prior work and it is no surprise that there are various publicly available evaluation datasets, e.g., SIGIR'05~\cite{Artiles2005}, WWW'05~\cite{Bekkerman2005}, WePS-2~\cite{Artiles2009}, etc. However, we found that available datasets are relatively small for conclusive statements on clustering quality. Furthermore, for the quality evaluation of our approach we needed a manual alignment of the search results with entity profiles from a given entity source. Such an alignment was not available in any of the datasets. Hence we decided to create a larger dataset, which would provide the required alignments. While an obvious application for our techniques is to use profile pages from social networks, e.g., LinkedIn, the general terms of agreement of those networks currently do not allow such usage. Therefore, we extracted  Wikipedia entities with ambiguous names; Web search results for those names were manually aligned with the corresponding articles (which we viewed as entity profiles). To this end we tested more than $85k$ possible entity -- Web page combinations for correctness.

\subsection{Dataset}
\label{sec:evaluation:data}
This subsection describes the content of the evaluation dataset, which we make available\footnote{\url{http://hpi-web.de/naumann/projekte/repeatability/datasets/wpsd.html}}.

\begin{figure}[ht]
 \centering
 \includegraphics[trim=55 275 55 315, clip=true, width=\linewidth]{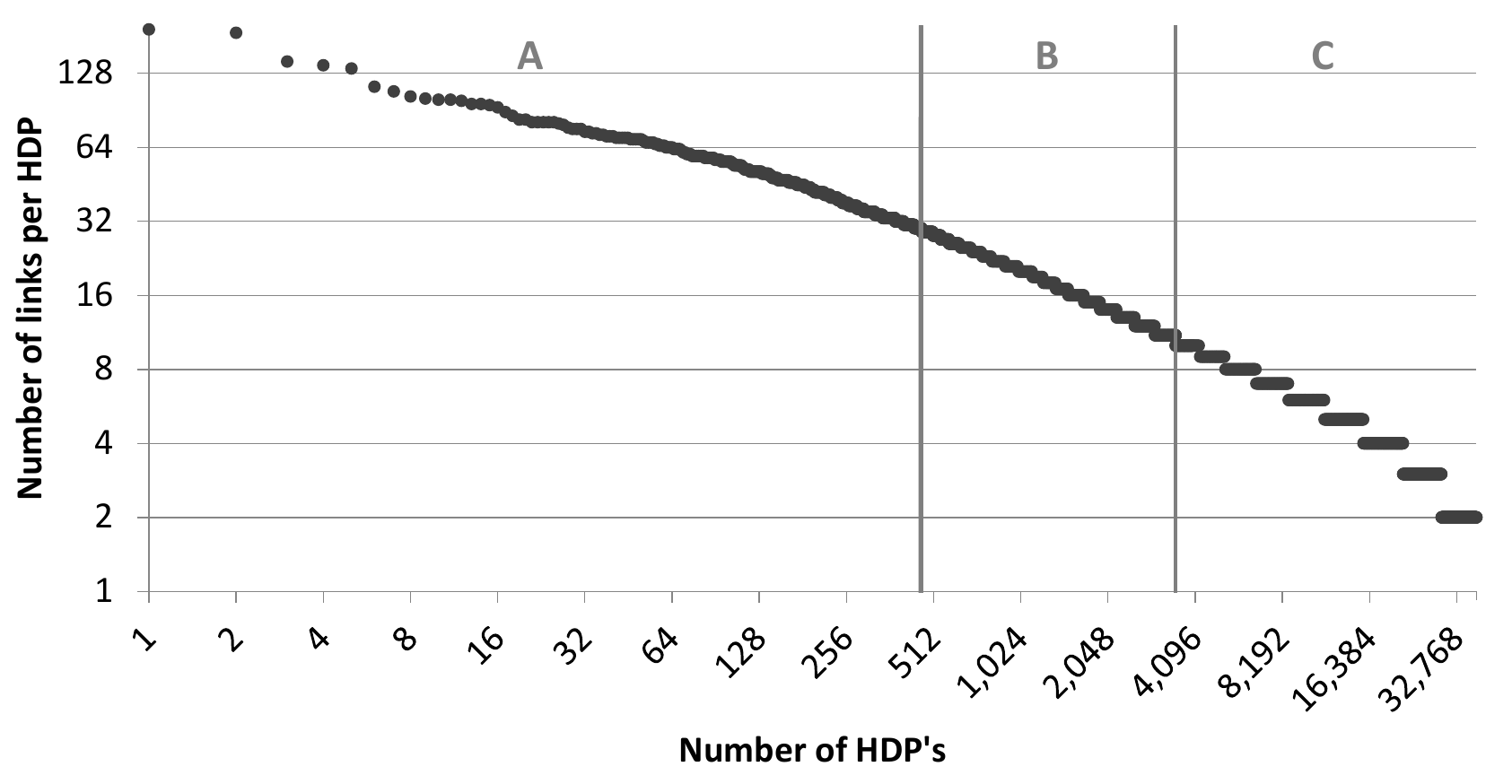}
 \caption{Distribution of articles linked from \WPDs on a log-log scale. The area under the curve is divided into three regions: (A) \WPDs containing highly ambiguous names with more than 24 linked articles each; this region contains $\sim 10\%$ of all article links in all \WPDs, (B) \WPDs containing ambiguous names with at least 10 linked articles each, corresponding to $\sim 23\%$ of all article links, and (C) \WPDs with moderately ambiguous names, containing $\sim 66\%$ of all article links.}
\label{fig:evaluation:ambigue_wiki_names}
\end{figure}

In \Wikipedia, articles about persons with ambiguous names are manually maintained and organized in so-called ``Human Name Disambiguation Pages'' (\WPDs for short). Each of these pages is dedicated to an ambiguous person name and contains links to \Wikipedia articles about persons of that name. Some \WPDs contain more links than others, reflecting the fact that on \Wikipedia some names are more ambiguous than others. Figure~\ref{fig:evaluation:ambigue_wiki_names} depicts the distribution of links over the \WPDs. As of July 2012 there were 38,692 \WPDs on \Wikipedia containing a total of 213,564 links referring to \Wikipedia articles. Interestingly, there were 28 \WPDs{} that did not contain any link (e.g.,\url{wiki:Sharon_Allen}) and 328 \WPDs contained only one valid link (e.g., see \url{wiki:Luciano_Buonaparte}). These articles were ignored. A careful, stratified random sampling procedure on the remaining 38,366 \WPDs, gave us the final dataset $\mathcal{X}$ with \WPDs sampled from three different regions of the distribution (i.e., A, B, C shown in Figure~\ref{fig:evaluation:ambigue_wiki_names}). The resulting sample contained \WPDs that were evenly spread across the three regions. In this way, a total of 50 \WPDs were extracted: 17 from region A (each with 24 linked articles or more), 17 from region B (containing 10+ linked articles), and 16 from the remaining long tail C of the distribution.

\begin{table}[ht]
\centering
\small
\begin{bigtabular}{|r|r|r|r|}
\hline
 & & &\\[-2ex]
\multicolumn{1}{|l|}{}                 & \multicolumn{1}{|l|}{\bf \#wiki}     & \multicolumn{1}{|l|}{\bf \#result} & \multicolumn{1}{|c|}{\bf alignments}                             \\
\multicolumn{1}{|l|}{\bf \WPD{} title} & \multicolumn{1}{|l|}{\bf articles} & \multicolumn{1}{|l|}{\bf pages}  & \multicolumn{1}{|l|}{\bf \#pages \hfill $\mapsto$ \hfill \#articles} \\
\hline
 & & &\\[-2ex]
\multicolumn{1}{|l|}{\bf A} & & &\\[-2.5ex]
John Campbell  & 100 & 96 & \hfill 33 \hfill $\mapsto$ \hfill 6 \hfill~ \\[-0.5ex]
James White  & 53 & 96 & \hfill 36 \hfill $\mapsto$ \hfill 6 \hfill~ \\[-0.5ex]
John Clark  & 42 & 96 & \hfill \phantom{0}4 \hfill $\mapsto$ \hfill 4 \hfill~ \\[-0.5ex]
John Rogers  & 42 & 93 & \hfill 28 \hfill $\mapsto$ \hfill 7 \hfill~ \\[-0.5ex]
William Murray  & 38 & 91 & \hfill 24 \hfill $\mapsto$ \hfill 8 \hfill~ \\[-0.5ex]
David Young  & 34 & 96 & \hfill 11 \hfill $\mapsto$ \hfill 4 \hfill~ \\[-0.5ex]
Robert Miller  & 34 & 97 & \hfill 16 \hfill $\mapsto$ \hfill 6 \hfill~ \\[-0.5ex]
James Baker (disam.)  & 33 & 97 & \hfill \phantom{0}3 \hfill $\mapsto$ \hfill 3 \hfill~ \\[-0.5ex]
John Mason  & 33 & 98 & \hfill 12 \hfill $\mapsto$ \hfill 5 \hfill~ \\[-0.5ex]
Thomas Baker  & 31 & 93 & \hfill \phantom{0}1 \hfill $\mapsto$ \hfill 1 \hfill~ \\[-0.5ex]
James Ross  & 30 & 96 & \hfill \phantom{0}7 \hfill $\mapsto$ \hfill 5 \hfill~ \\[-0.5ex]
Michael Kelly  & 29 & 92 & \hfill 14 \hfill $\mapsto$ \hfill 2 \hfill~ \\[-0.5ex]
David Thompson  & 28 & 97 & \hfill 26 \hfill $\mapsto$ \hfill 5 \hfill~ \\[-0.5ex]
James Gordon  & 28 & 90 & \hfill 31 \hfill $\mapsto$ \hfill 4 \hfill~ \\[-0.5ex]
John Richards  & 26 & 94 & \hfill \phantom{0}0 \hfill $\mapsto$ \hfill 0 \hfill~ \\[-0.5ex]
Thomas Bell  & 26 & 89 & \hfill \phantom{0}6 \hfill $\mapsto$ \hfill 5 \hfill~ \\[-0.5ex]
William Collins  & 24 & 87 & \hfill 27 \hfill $\mapsto$ \hfill 5 \hfill~ \\
\hline
 & & &\\[-2ex]
\multicolumn{1}{|l|}{\bf B} & & & \\[-2.5ex]
~~~~Maria Theresa (disam.)  & 19 & 91 & \hfill 31 \hfill $\mapsto$ \hfill 1 \hfill~ \\[-0.5ex]
Mark Robinson  & 16 & 94 & \hfill \phantom{0}6 \hfill $\mapsto$ \hfill 3 \hfill~ \\[-0.5ex]
Michael Walsh  & 16 & 93 & \hfill 44 \hfill $\mapsto$ \hfill 4 \hfill~ \\[-0.5ex]
Danny Williams  & 13 & 92 & \hfill 37 \hfill $\mapsto$ \hfill 4 \hfill~ \\[-0.5ex]
David Brooks  & 13 & 97 & \hfill 86 \hfill $\mapsto$ \hfill 1 \hfill~ \\[-0.5ex]
Fred Johnson  & 13 & 84 & \hfill \phantom{0}4 \hfill $\mapsto$ \hfill 2 \hfill~ \\[-0.5ex]
George Stephens  & 13 & 90 & \hfill \phantom{0}1 \hfill $\mapsto$ \hfill 1 \hfill~ \\[-0.5ex]
Gustav of Sweden  & 13 & 81 & \hfill 27 \hfill $\mapsto$ \hfill 6 \hfill~ \\[-0.5ex]
Henry Robinson  & 13 & 95 & \hfill \phantom{0}1 \hfill $\mapsto$ \hfill 1 \hfill~ \\[-0.5ex]
John MacArthur  & 13 & 99 & \hfill 97 \hfill $\mapsto$ \hfill 5 \hfill~ \\[-0.5ex]
Billy Hayes  & 11 & 91 & \hfill 49 \hfill $\mapsto$ \hfill 4 \hfill~ \\[-0.5ex]
Drogo  & 11 & 93 & \hfill 77 \hfill $\mapsto$ \hfill 5 \hfill~ \\[-0.5ex]
Kimberley Smith  & 11 & 97 & \hfill 14 \hfill $\mapsto$ \hfill 4 \hfill~ \\[-0.5ex]
William Hammond  & 11 & 95 & \hfill 20 \hfill $\mapsto$ \hfill 6 \hfill~ \\[-0.5ex]
Bob Hope (disam.)  & 10 & 90 & \hfill 78 \hfill $\mapsto$ \hfill 5 \hfill~ \\[-0.5ex]
John Penn  & 10 & 95 & \hfill 31 \hfill $\mapsto$ \hfill 4 \hfill~ \\[-0.5ex]
Robert Dean  & 10 & 92 & \hfill 20 \hfill $\mapsto$ \hfill 2 \hfill~ \\
\hline
 & & &\\[-2ex]
\multicolumn{1}{|l|}{\bf C} & & &\\[-2.5ex]
Michael Myers  & 8 & 74 & \hfill 44 \hfill $\mapsto$ \hfill 3 \hfill~ \\[-0.5ex]
Edward Cole  & 6 & 94 & \hfill \phantom{0}8 \hfill $\mapsto$ \hfill 2 \hfill~ \\[-0.5ex]
Peter Donnelly (disam.)  & 6 & 98 & \hfill 31 \hfill $\mapsto$ \hfill 5 \hfill~ \\[-0.5ex]
Siemowit of Masovia  & 5 & 99 & \hfill 10 \hfill $\mapsto$ \hfill 3 \hfill~ \\[-0.5ex]
William McKee  & 5 & 98 & \hfill \phantom{0}2 \hfill $\mapsto$ \hfill 1 \hfill~ \\[-0.5ex]
Brian Reynolds  & 4 & 95 & \hfill 14 \hfill $\mapsto$ \hfill 1 \hfill~ \\[-0.5ex]
Leslie Baker  & 4 & 93 & \hfill \phantom{0}9 \hfill $\mapsto$ \hfill 2 \hfill~ \\[-0.5ex]
George Donnelly  & 3 & 99 & \hfill \phantom{0}4 \hfill $\mapsto$ \hfill 2 \hfill~ \\[-0.5ex]
Peter Robertson  & 3 & 97 & \hfill \phantom{0}2 \hfill $\mapsto$ \hfill 1 \hfill~ \\[-0.5ex]
Timothy Allen (disam.)  & 3 & 92 & \hfill 29 \hfill $\mapsto$ \hfill 2 \hfill~ \\[-0.5ex]
Daniel Terra  & 2 & 100 & \hfill 15 \hfill $\mapsto$ \hfill 1 \hfill~ \\[-0.5ex]
Frida Svensson  & 2 & 98 & \hfill 54 \hfill $\mapsto$ \hfill 1 \hfill~ \\[-0.5ex]
Jack Renner  & 2 & 95 & \hfill 24 \hfill $\mapsto$ \hfill 2 \hfill~ \\[-0.5ex]
Lasse Nielsen  & 2 & 91 & \hfill \phantom{0}8 \hfill $\mapsto$ \hfill 2 \hfill~ \\[-0.5ex]
William McFaddin  & 2 & 96 & \hfill \phantom{0}8 \hfill $\mapsto$ \hfill 2 \hfill~ \\[-0.5ex]
William Nicol  & 2 & 97 & \hfill \phantom{0}3 \hfill $\mapsto$ \hfill 1 \hfill~ \\
\hline
\end{bigtabular}
\caption{Overview of the manually labeled evaluation dataset.
}
\label{tab:evaluation:sample_ambigue_names}
\end{table}

Given an \WPD from the sample ($\mathcal{X}$), the ambiguous person name $x$ can be directly inferred from the url, e.g., \url{wiki:John_Campbell} is translated to $x = $``John Campbell'', ignoring \Wikipedia notations, such as `\texttt{\_(disambiguation)}'. Additionally, the following five steps where performed:
\begin{enumerate*}
\item The content of the \WPD was extracted using \Wikipedia's MediaWiki RESTful API\footnote{\url{http://en.wikipedia.org/w/api.php}}.
\item The set $E$ of \Wikipedia articles referenced from the \WPD was extracted.
\item The textual content of each article in $E$ was parsed, tokenized, and cleaned (from any markups) to derive unigram features.
\item A Google Custom Search (using the Google API\footnote{\url{http://developers.google.com/custom-search/}}) was performed for the person name $x$ derived from the \WPDs url, whereas \Wikipedia articles in the results were excluded, e.g., for $x=$``John Campbell'' the Google query ``John Campbell -site:wikipedia.org'' was issued.
\item The top-100 results were retrieved (ignoring empty, non-text\-ual, and HTTP error pages),
e.g.:
\begin{small}
\begin{list}{$\bullet$}
  { \setlength{\itemsep}{0pt}
    \setlength{\parsep}{0pt}
    \setlength{\topsep}{1pt}
    \setlength{\partopsep}{0pt}
    \setlength{\partopsep}{0pt}
    \setlength{\leftmargin}{1.5em}
    \setlength{\labelwidth}{1.3em}
    \setlength{\labelsep}{0.2em} }
\item[(1)] \url{http://yelp.com/biz/john-campbells-irish-}\\\url{bakery-san-francisco-4}
\item[(2)] \url{http://www.facebook.com/pages/John-Campbell-}\\\url{Yoga/125627657452140}
\item[(3)] \url{http://economics.ag.utk.edu/campbell.html}
\item[(4)] \url{http://campbell.house.gov/index.php?id=1036&}\\\url{option=com_content}
\item[\vdots]
\end{list}
\end{small}
Subsequently the same procedure as in Step 3 was applied to derive features from each of the documents in the result set.
\end{enumerate*}

To create the gold standard for the evaluation, for each query the alignment of search results with \Wikipedia articles was carefully performed by human labelers, namely students from our department. Each document in the result set was either assigned to exactly one \Wikipedia article (referenced from the \WPD from which the query was derived) or labeled as noise document if it was not about any of the entities described by the \Wikipedia articles. Also, in cases in which multiple \Wikipedia entities occurred in the result document, the document was labeled as noise document. This process added up to more than $85k$ possible Web page to entity (non-noise) combinations (i.e., $\sum_{x} |E_x| \times |D_x|$) that had to be checked manually.

Table~\ref{tab:evaluation:sample_ambigue_names} gives an overview of the manually aligned dataset. One alignment task is identified by the \textbf{\WPD title}, the \Wikipedia articles referenced from the \WPD (\textbf{\#wiki articles}), and the remaining Web pages from Google's top-100 results after excluding empty, non-textual, and HTTP-error pages (\textbf{\#result pages}). The rightmost column in Table~\ref{tab:evaluation:sample_ambigue_names} (\textbf{alignments}) shows the number of result pages (\textbf{\#pages}) that could be assigned to any \Wikipedia article referenced from the \WPD and the number of different \Wikipedia articles (\textbf{\#articles}), i.e., ground-truth classes, to which result pages were assigned. The remaining search results that could not be aligned with any \Wikipedia article referenced from the \WPD as well as the ones that were about multiple individuals were assigned to a designated class \textNoiseProfile.

A common but misleading assumption that is based on the notability of \Wikipedia entities is that searching for \Wikipedia entity names yields many top results related to the corresponding \Wikipedia entities. This assumption holds when the documents about \Wikipedia entities are also popular on the Web/Google (which is often the case), but for niche \Wikipedia entities, which are known to few scholars, this assumption leads astray. For instance, ``John Campbell'' refers to 100 different individuals in \Wikipedia, but only 6 of them actually occurred in Google's top-100 results (after excluding Wikipdia-related results). In fact, this skew is the case for the majority of the ambiguous names in our dataset. Also note that the classification problem is extremely difficult: for the ambiguous name ``John Campbell'' the problem is to automatically classify 96 Web pages into one of 100 entities ($+\noiseProfile{}$), while the ground truth tells us that only 33 out of 96 results are assigned to 6 out of 100 \Wikipedia articles and the rest to \textNoiseProfile. A more extreme example for this observation is ``John Richards'', for which no valid mapping between the top-100 search results and any of \Wikipedia articles referenced from the corresponding \WPD was found. Hence using \Wikipedia to bootstrap the grouping of search results to ambiguous person names can be quite challenging and also covers the problem of clustering documents about less famous people on the Web. As stated earlier, the entities of \textNoiseProfile are not classified with respect to the entity source and are thus unclustered.

\subsection{Comparison with clustering techniques}
\label{sec:evaluation:results:clustering}
This section provides a comparison of state-of-the-art clustering algorithms to our techniques. The results demonstrate the advantages of the proposed bootstrapping approach, which exploits prior knowledge to perform the disambiguation task.

For the comparison, we selected two of the most popular clustering methods: Hierarchical Agglomerative Clustering (HAC), and \textKMeans, as provided by the Weka machine learning toolkit\footnote{\url{http://www.cs.waikato.ac.nz/ml/weka/}}. For each of the methods, we tested many different configurations. For example, for fairness reasons, one of the configurations of \textKMeans used the \Wikipedia entities as initial centroids similarly as proposed in~\cite{Berendsen2012}. However, random seeds\footnote{We used the average results over ten repetitions.} led to better results than the above configuration.

For the HAC method we tried different linkage configurations, e.g., single-link, average-link, complete-link~\cite{Manning2008}. The configuration with the complete linkage criterion worked best and was chosen as a competitor. For both methods we tested configurations in which the parameter $k$ of resulting clusters was set to the number of different \Wikipedia entities for the corresponding ambiguous name.

For comparison with the above methods we used an implementation of \textSmoothedScoringFunction and \textMultinomialJelinekProb. Both methods were applied using the intersection-noise entity (\textIntersectionNoiseProfile) to derive a mapping as proposed in Section~\ref{sec:algorithm}. The mapping results were transformed to anonymous clusters by omitting the assigned \Wikipedia entity label.

To be fair, for the quality evaluation of the groups returned by the clustering methods, result documents that were not related to any of the \Wikipedia entities (these would be noise documents for our bootstrapping approach) were not taken into account. The reason is that the clustering algorithms treat such documents equally to all the others, thus missing the task of creating a coherent ``noise cluster'' (i.e., with documents assigned to \textNoiseProfile). This led to 1,095 Web documents used for the clustering evaluation.

To compare the clustering performance of the approaches we applied two established measures: purity and normalized mutual information (NMI)\cite{Manning2008}. Table~\ref{tab:evaluation:clustering_relevant} shows the average purity and NMI values over all 50~names from $\mathcal{X}$. As can be seen, all three bootstrapping-based classification models outperform  the clustering approaches with respect to both measures. However, the differences for the purity values are smaller than for the NMI values.

\begin{table}[ht]
\centering
\begin{bigtabular}{|r|r|r|}
\hline& & \\[-2.2ex]
           &     {\bf purity} &        {\bf NMI} \\
\hline& & \\[-2.2ex]

{\boldmath$\smoothedScoringFunction$\unboldmath}  &     {\bf 0.913 } &     {\bf 0.560} \\

{\boldmath$\multinomialJelinekProb$\unboldmath} &      0.890 &      0.492 \\

\hline& & \\[-2.2ex]
{\bf HAC} &      0.829 &      0.321 \\

{\bf \textKMeans} &      0.814 &      0.287 \\
\hline
\end{bigtabular}

\caption{Clustering evaluation of documents related to a \Wikipedia article.}
\label{tab:evaluation:clustering_relevant}
\end{table}

The purity is a commonly used evaluation measure in the area of personal name resolution. However, a high purity is easy to achieve with a large number of clusters (i.e., a score of~1 can be achieved by turning every document into one cluster). The results for the clustering methods align with the evaluation of Balog et al.~\cite{Balog08}. This shows that even the subset of our evaluation dataset (i.e., Web pages related to \Wikipedia entities) is representative for the evaluation of a personal name resolution task.

The NMI measure shows larger deviations, because it is normalized by the overall entropy across clusters (which requires clusters not only to possibly contain elements from only one class, but also to possibly contain all the elements from that class). In terms of NMI score, all two bootstrapping approaches outperform the unsupervised methods by a large margin. This shows that our approach is able to balance between quality and size of clusters.

However, note that although this evaluation shows promising results for the grouping of this subset of documents, the actual problem covered in this work also depends on another difficult subtask, namely the identification of documents not related to any entity of the bootstrapping source. The following section discusses the evaluation of our approaches for the task of linking Web documents to ambiguous Wikipedia entities and thus covers the complete problem.

\subsection{Probabilistic vs. vector space models}
\label{sec:evaluation:results}
\begin{figure*}[t!]
\begin{center}$
\begin{array}{cccc}
 &
\parbox{55pt}{\hfill\text{{\bf no noise}}} & \text{{\bf union-noise}} & \text{{\bf intersection-noise}} \\

\begin{sideways}\parbox{66pt}{\hfill\text{\textbf{(a)} {\boldmath$\similarityScoringFunction$\unboldmath}}}\end{sideways} & \includegraphics[trim=60 310 60 310, clip=true, width=166.5pt]{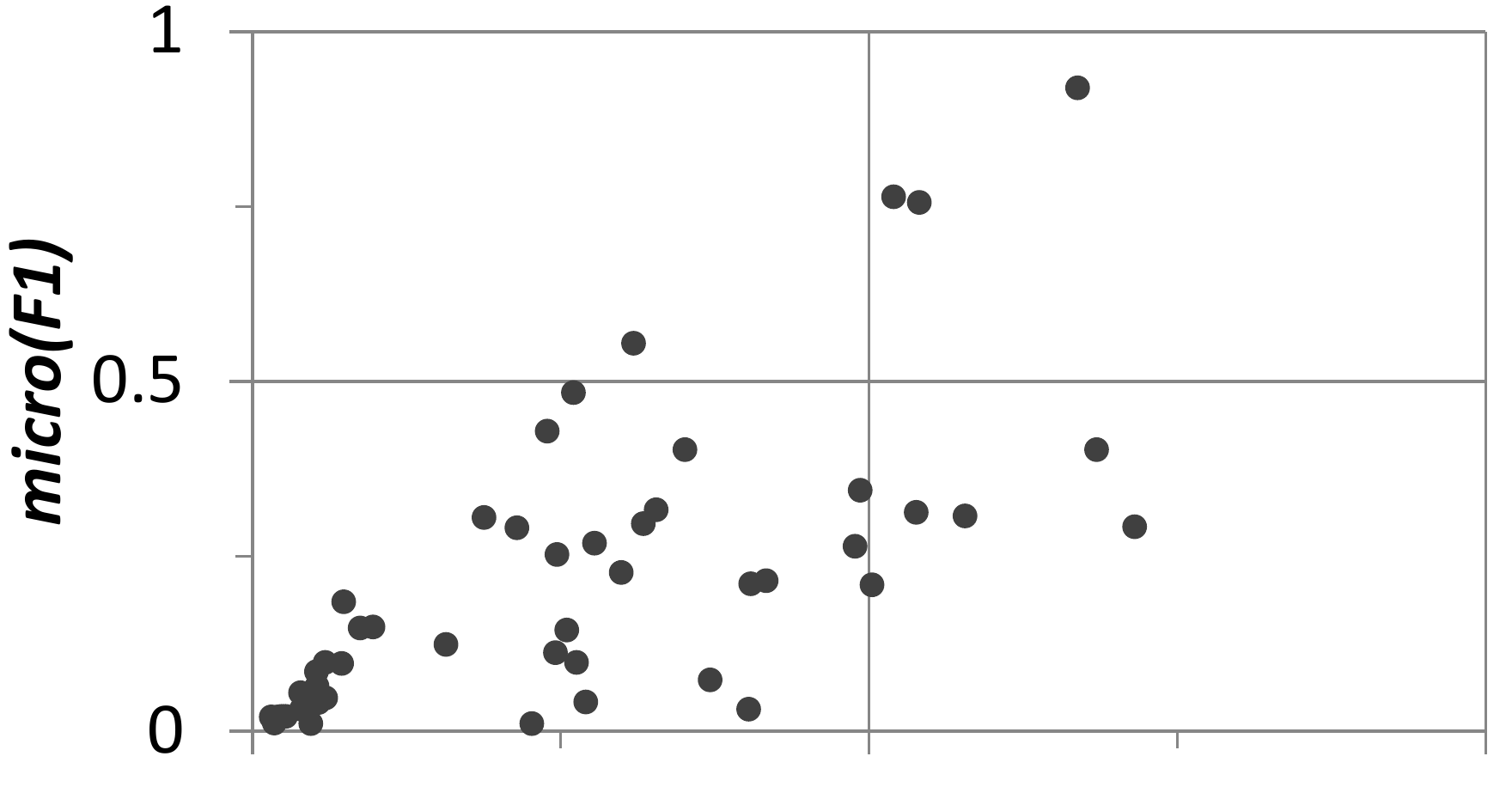} & \includegraphics[trim=130 310 60 310, clip=true, width=143pt]{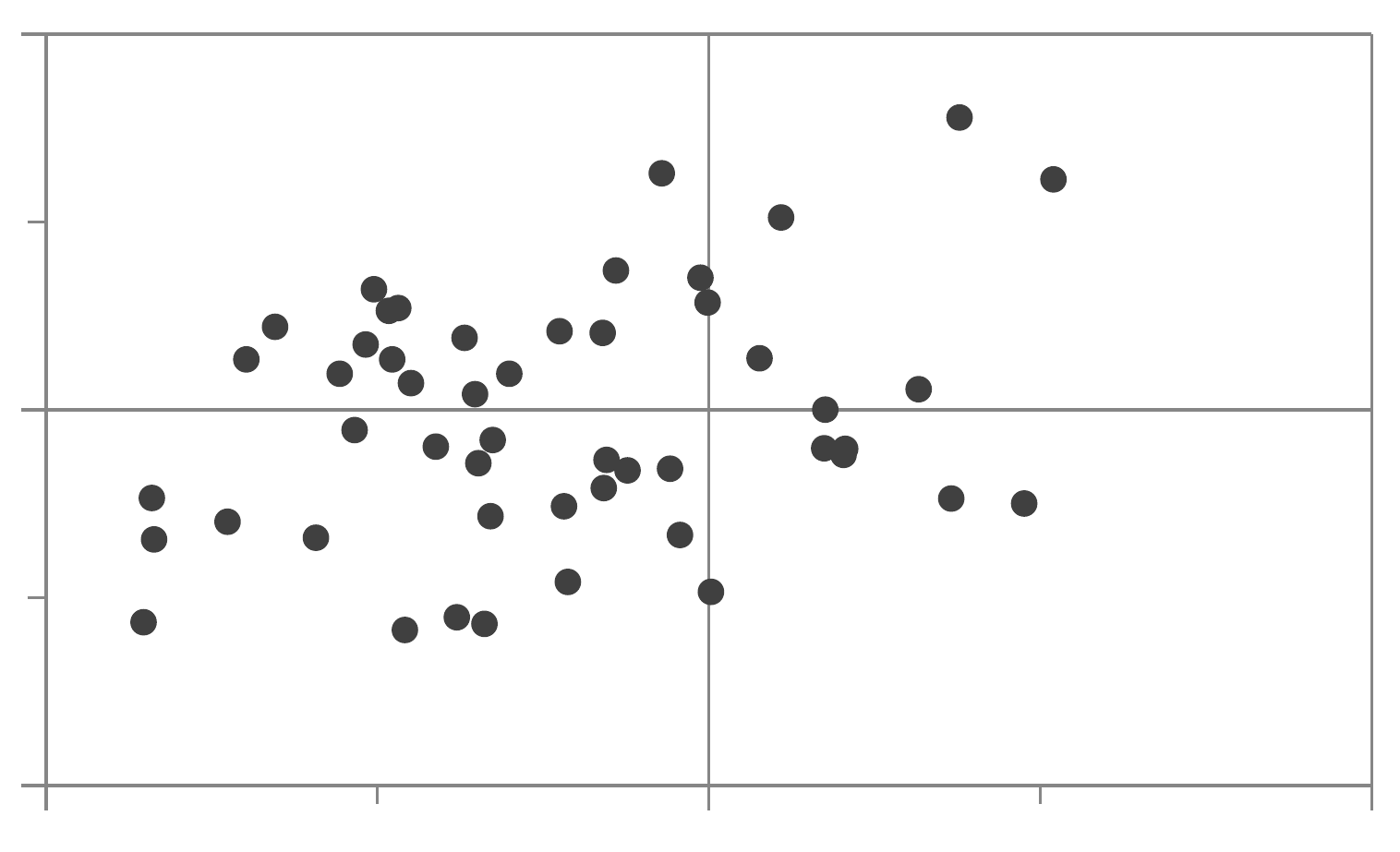} & \includegraphics[trim=130 310 60 310, clip=true, width=143pt]{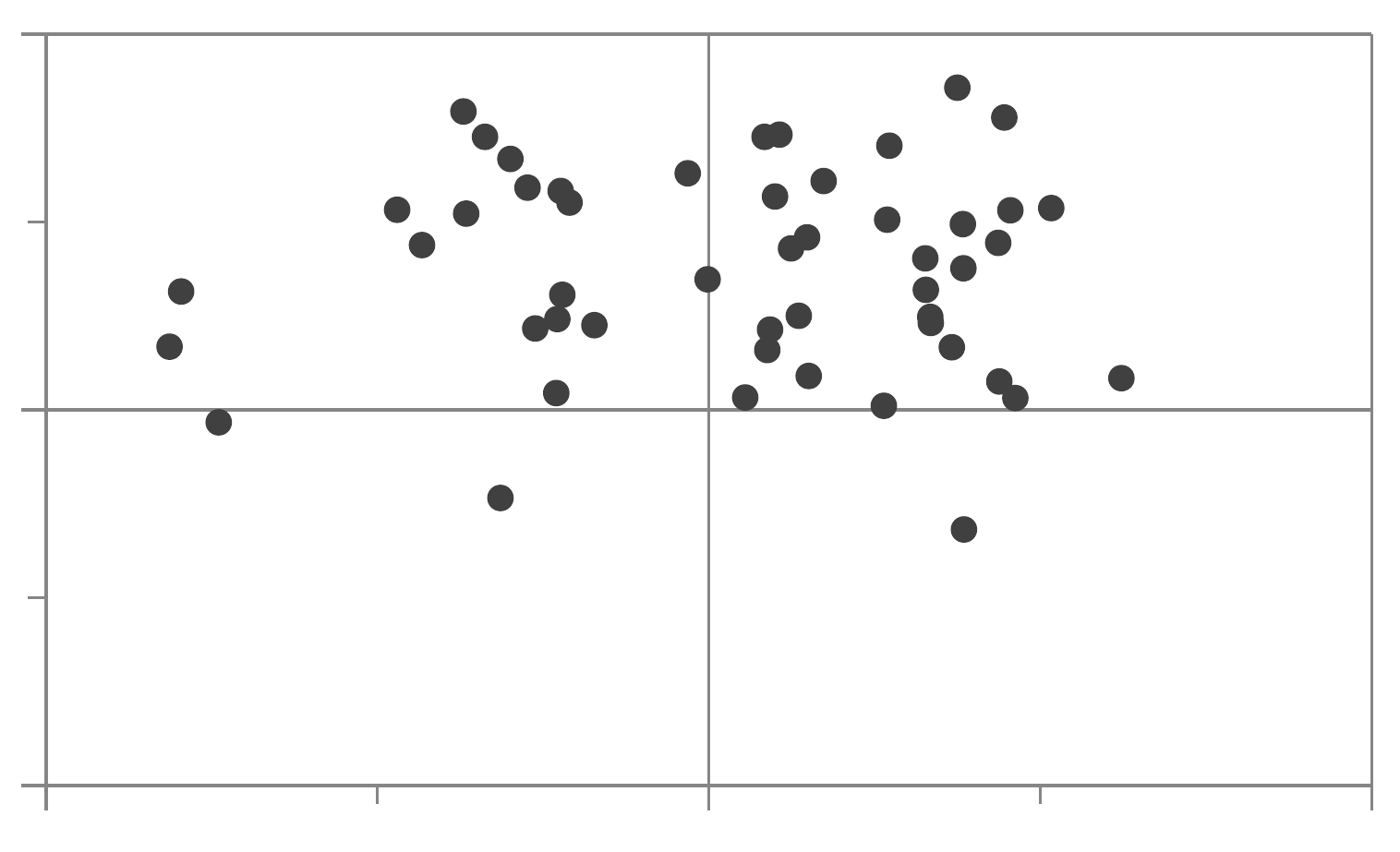} 
\\
\\
\begin{sideways}\parbox{64pt}{\hfill\text{\textbf{(b)} {\boldmath$\scoringFunction$\unboldmath}}}\end{sideways} &
\includegraphics[trim=60 310 60 310, clip=true, width=166.5pt]{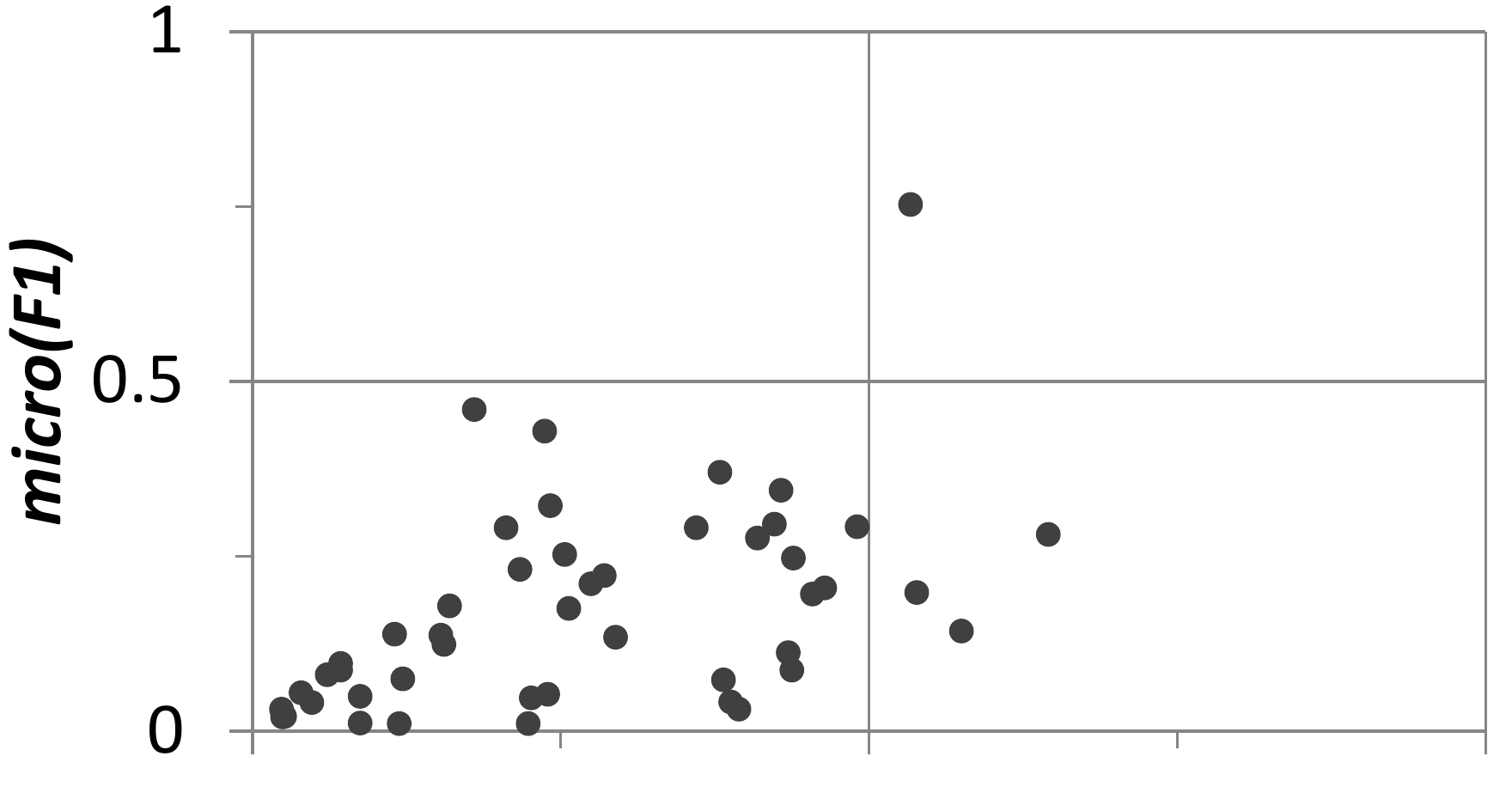} & \includegraphics[trim=130 310 60 310, clip=true, width=143pt]{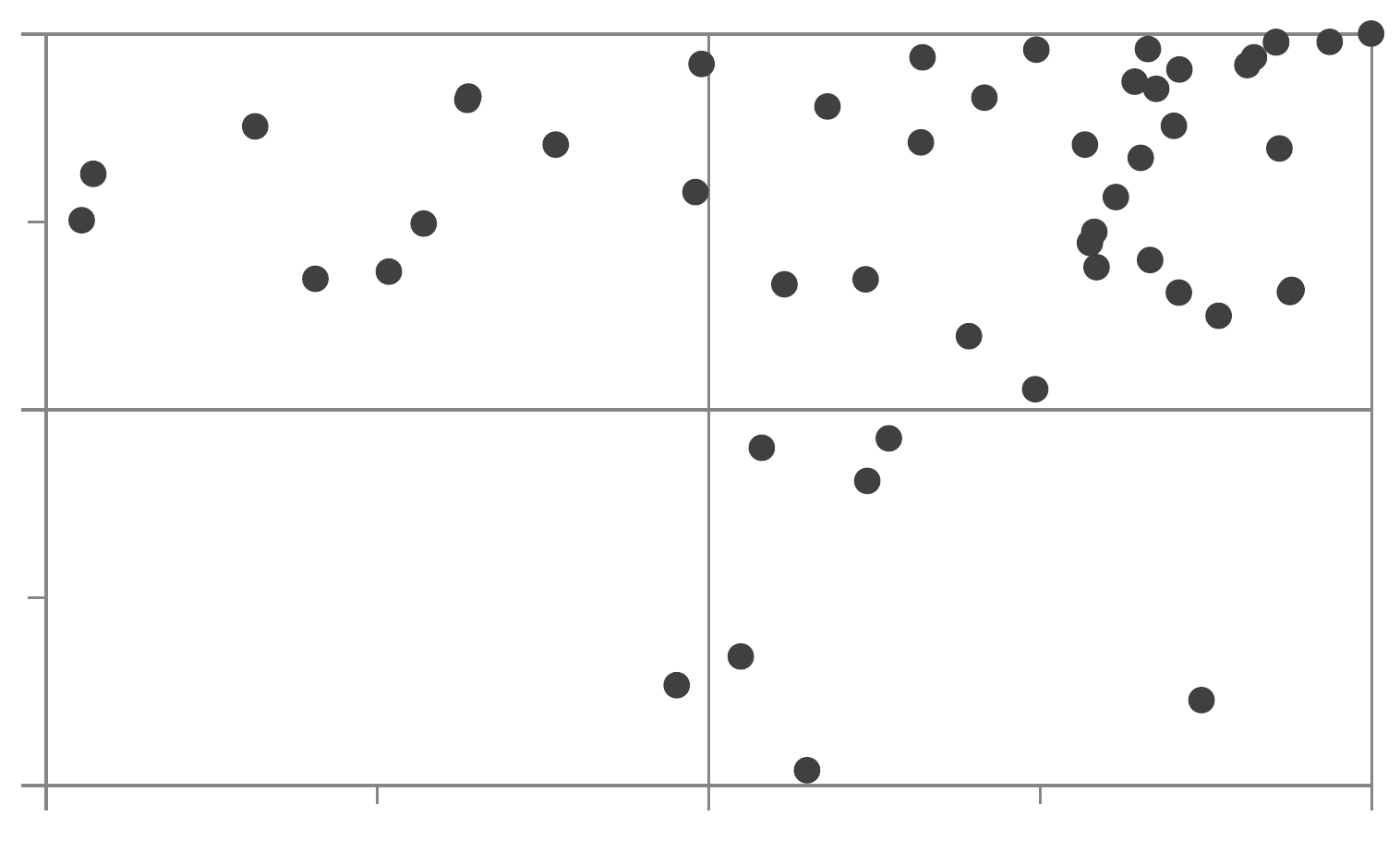} & \includegraphics[trim=130 310 60 310, clip=true, width=143pt]{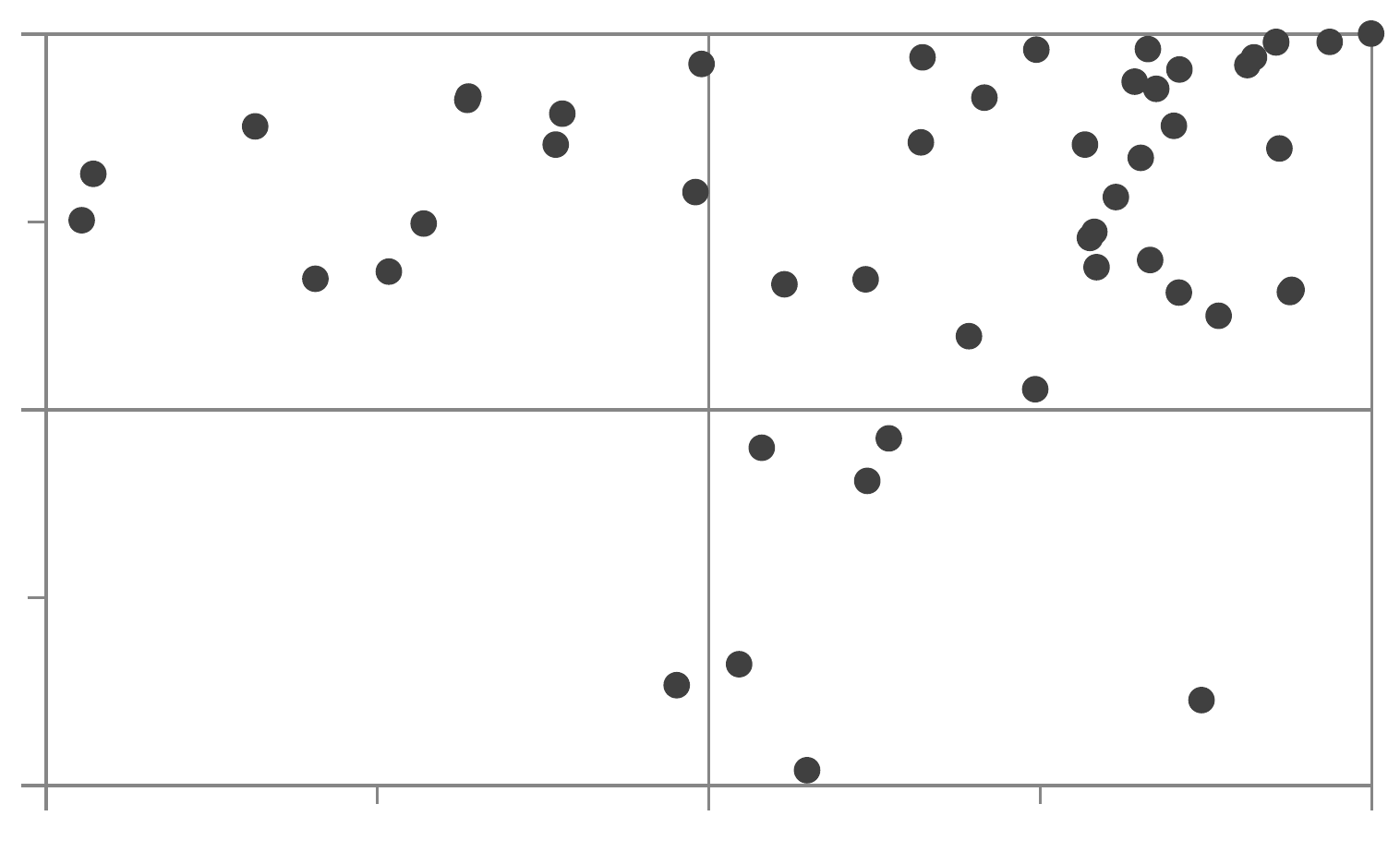} 
\\
\\
\begin{sideways}\parbox{66pt}{\hfill\text{\textbf{(c)} {\boldmath$\smoothedScoringFunction$\unboldmath}}}\end{sideways} &
\includegraphics[trim=60 310 60 310, clip=true, width=166.5pt]{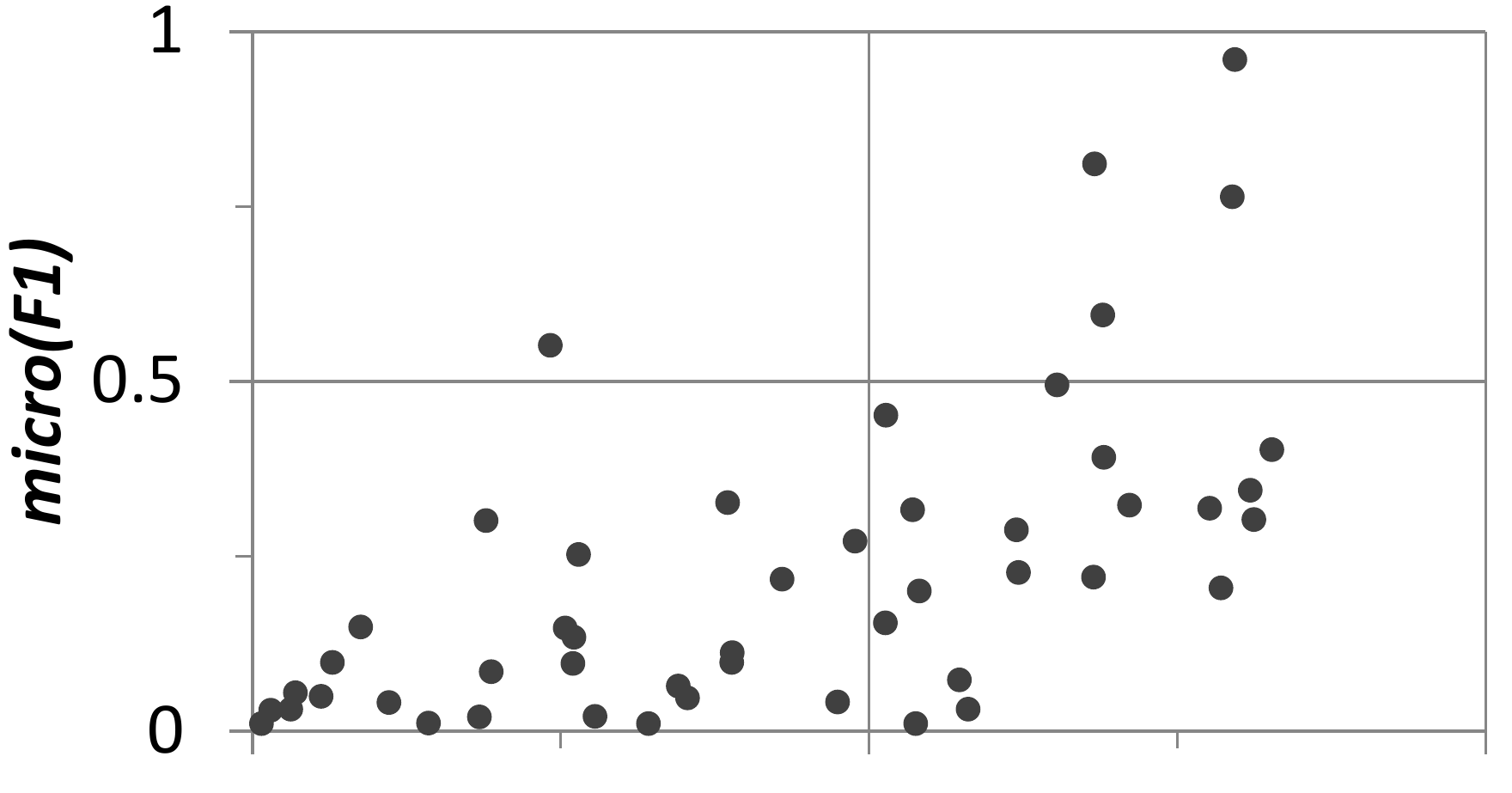} & \includegraphics[trim=130 310 60 310, clip=true, width=143pt]{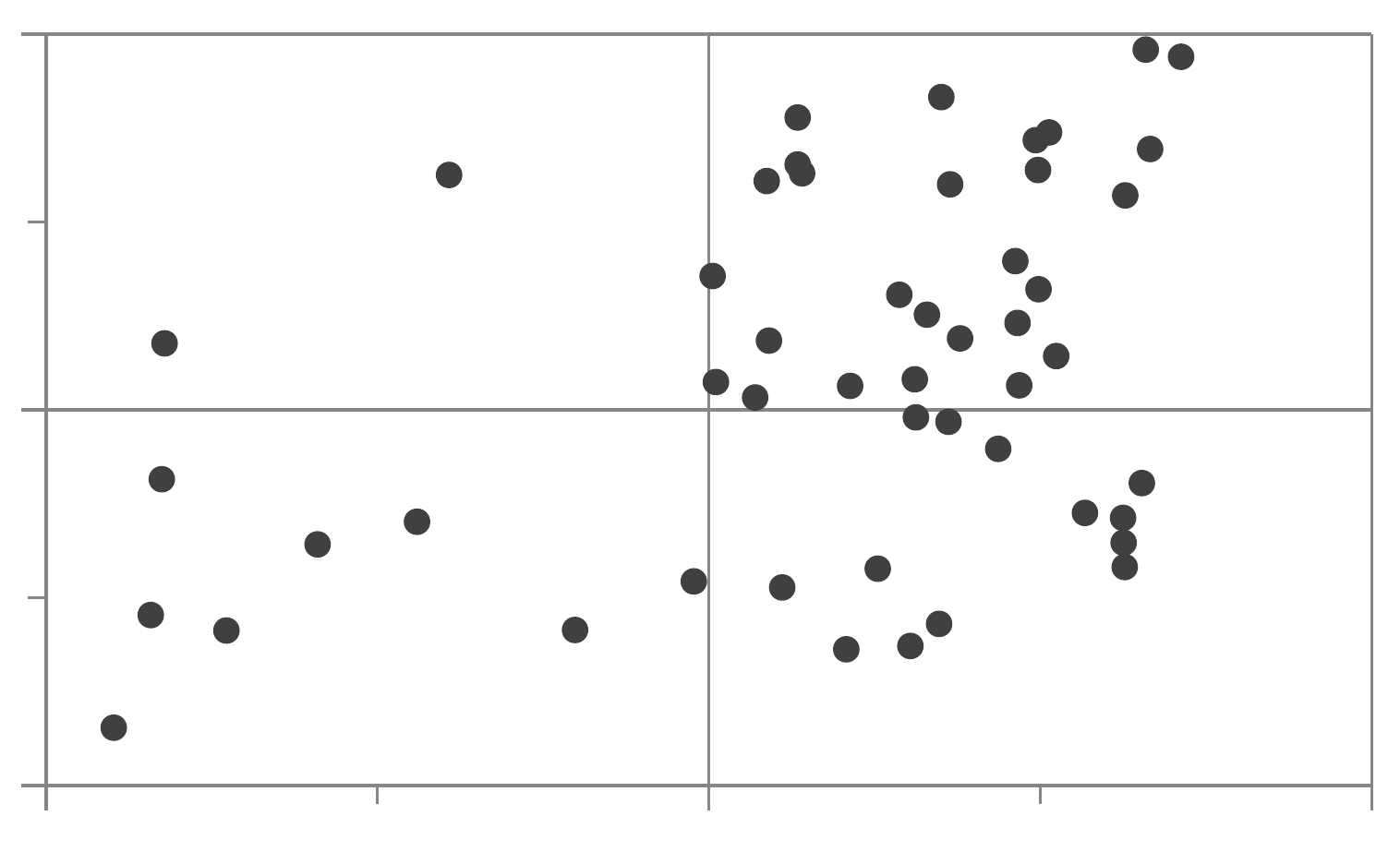} & \includegraphics[trim=130 310 60 310, clip=true, width=143pt]{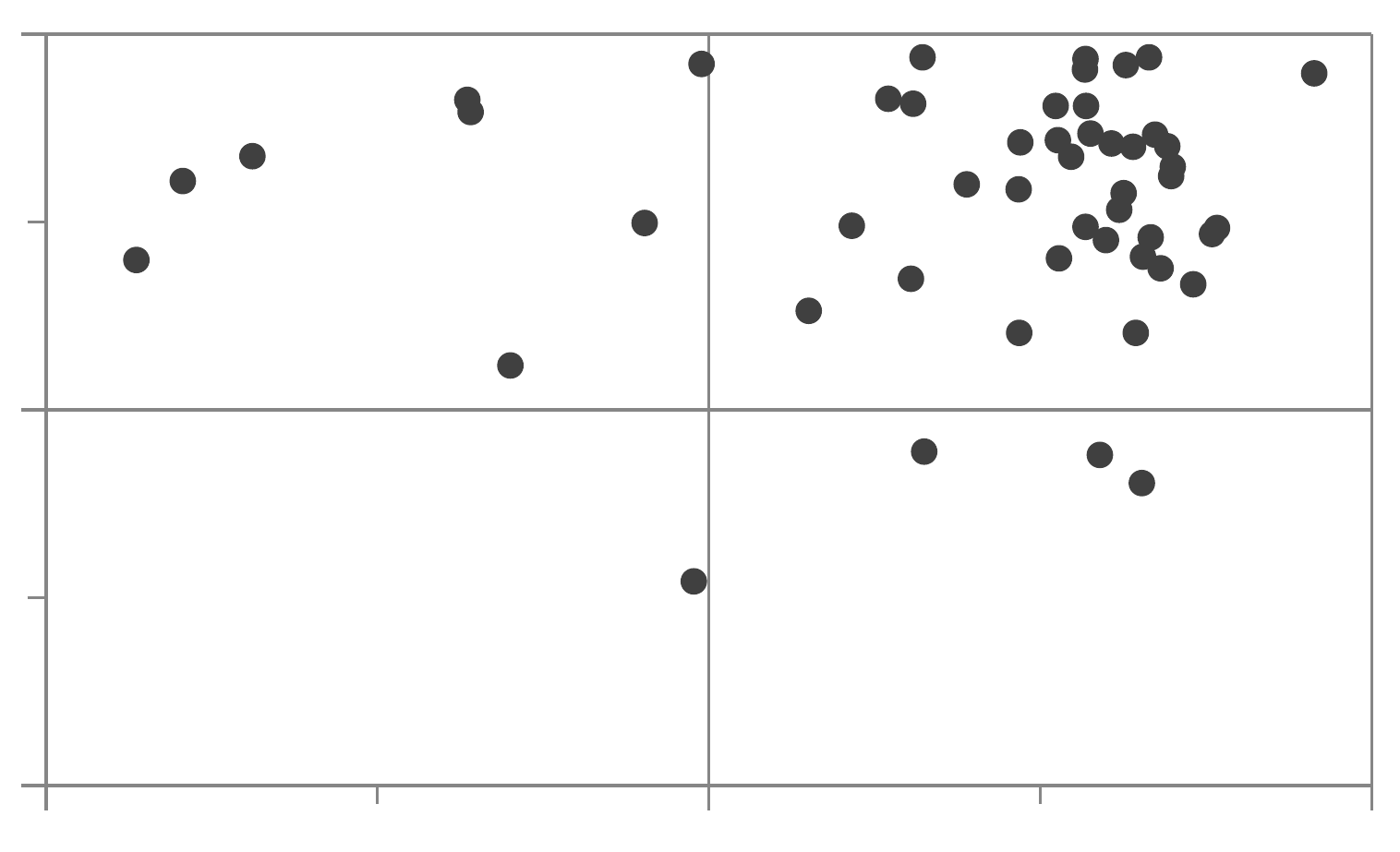} 
\\
\\
\begin{sideways}\parbox{57pt}{\hfill\text{\textbf{(d)} {\boldmath$\bernoulliLaplaceProb$\unboldmath}}}\end{sideways} &
\includegraphics[trim=60 310 60 310, clip=true, width=166.5pt]{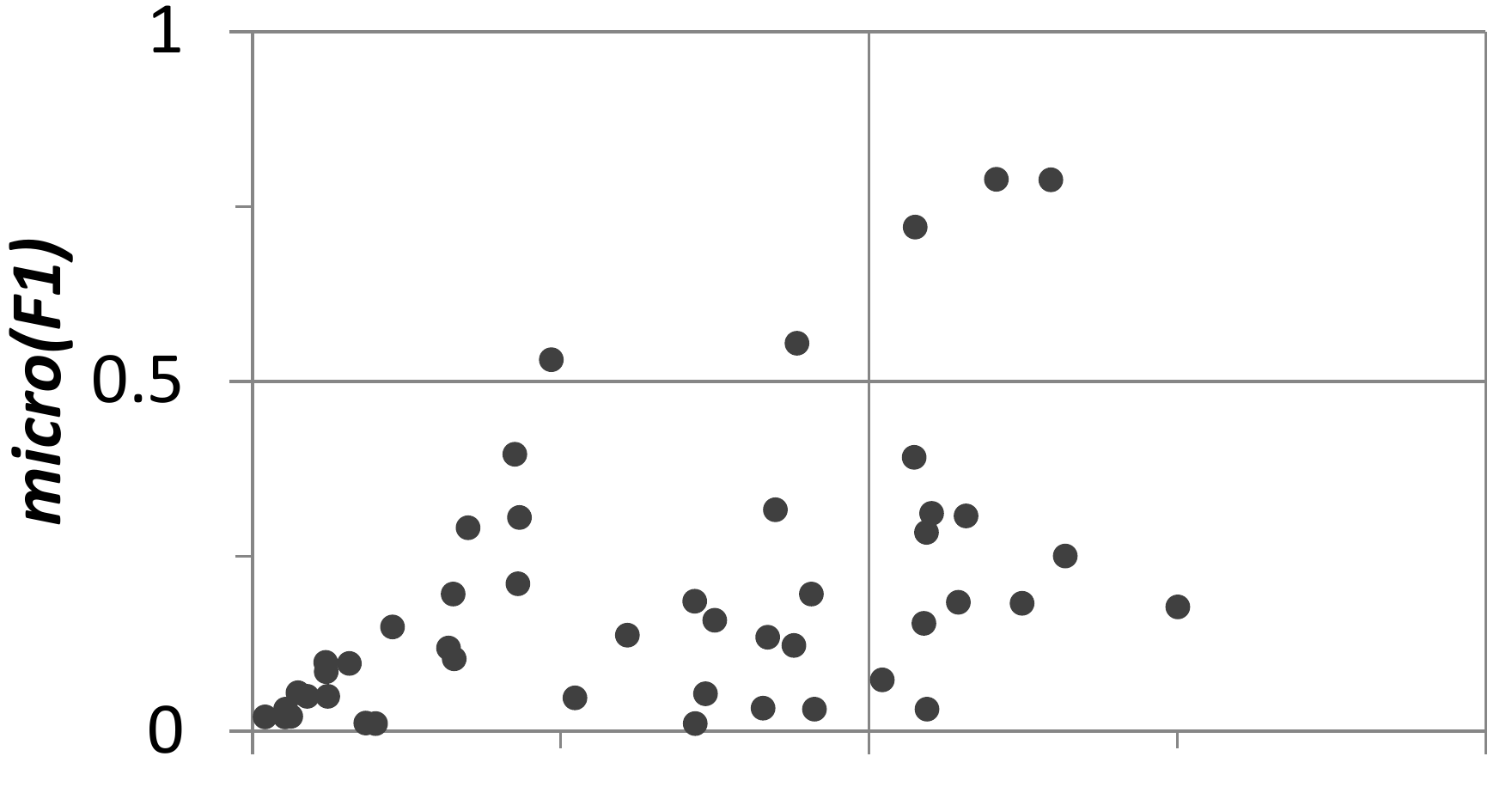} & \includegraphics[trim=130 310 60 310, clip=true, width=143pt]{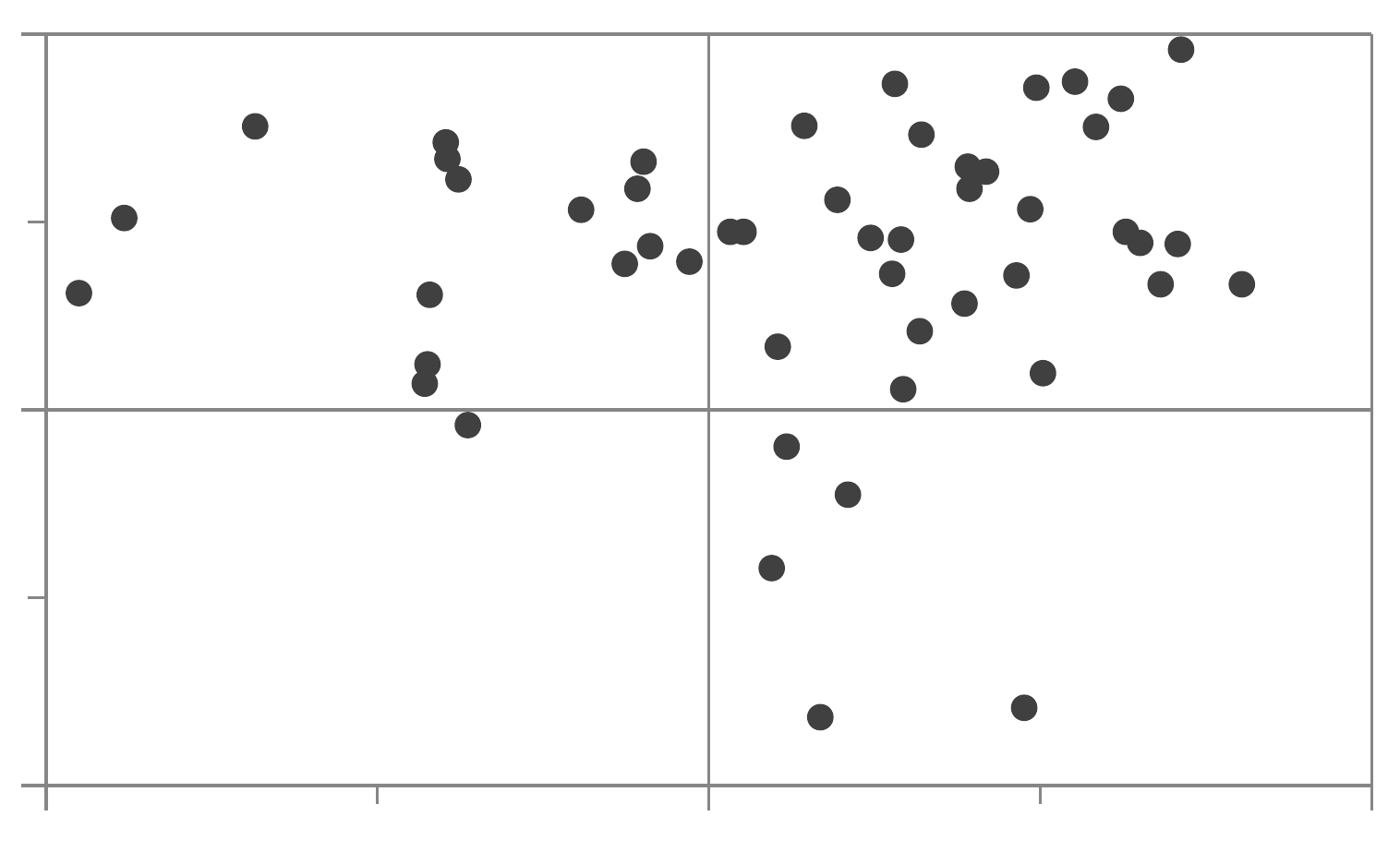} & \includegraphics[trim=130 310 60 310, clip=true, width=143pt]{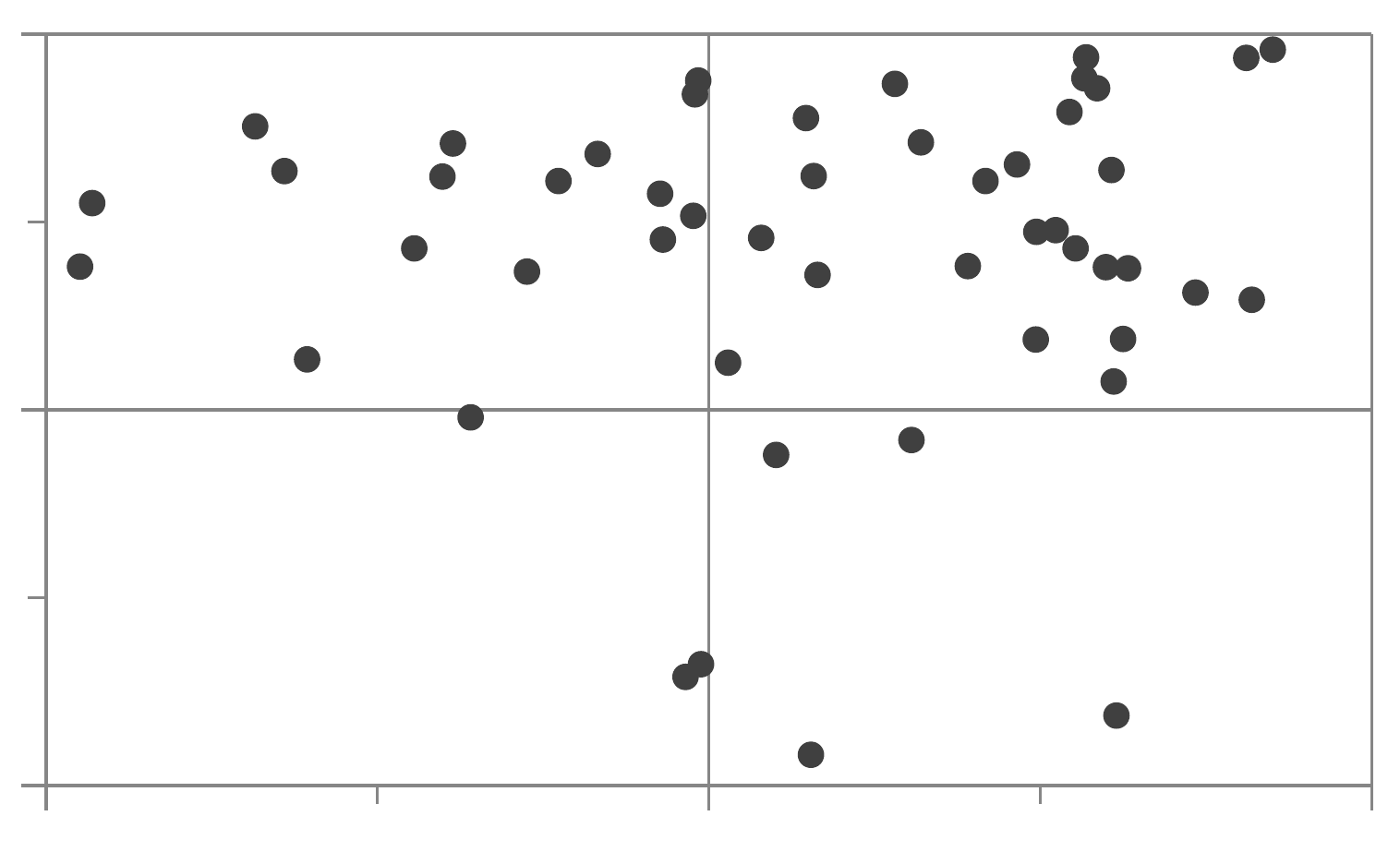} 
\\
\\
\begin{sideways}\parbox{75pt}{\hfill\text{\textbf{(e)} {\boldmath$\multinomialJelinekProb$\unboldmath}}}\end{sideways} &
\includegraphics[trim=60 265 60 310, clip=true, width=166.5pt]{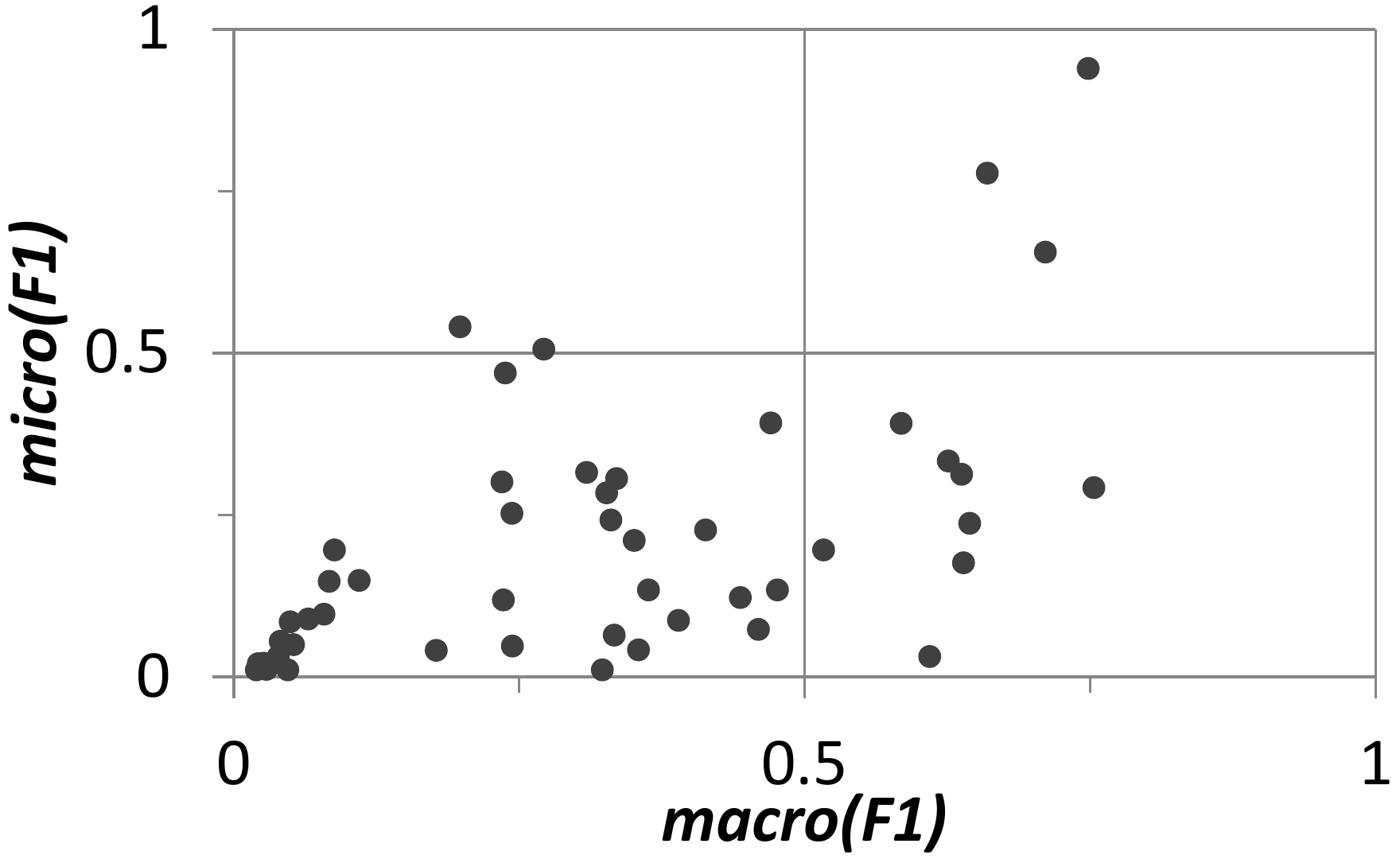} & \includegraphics[trim=130 265 60 310, clip=true, width=143pt]{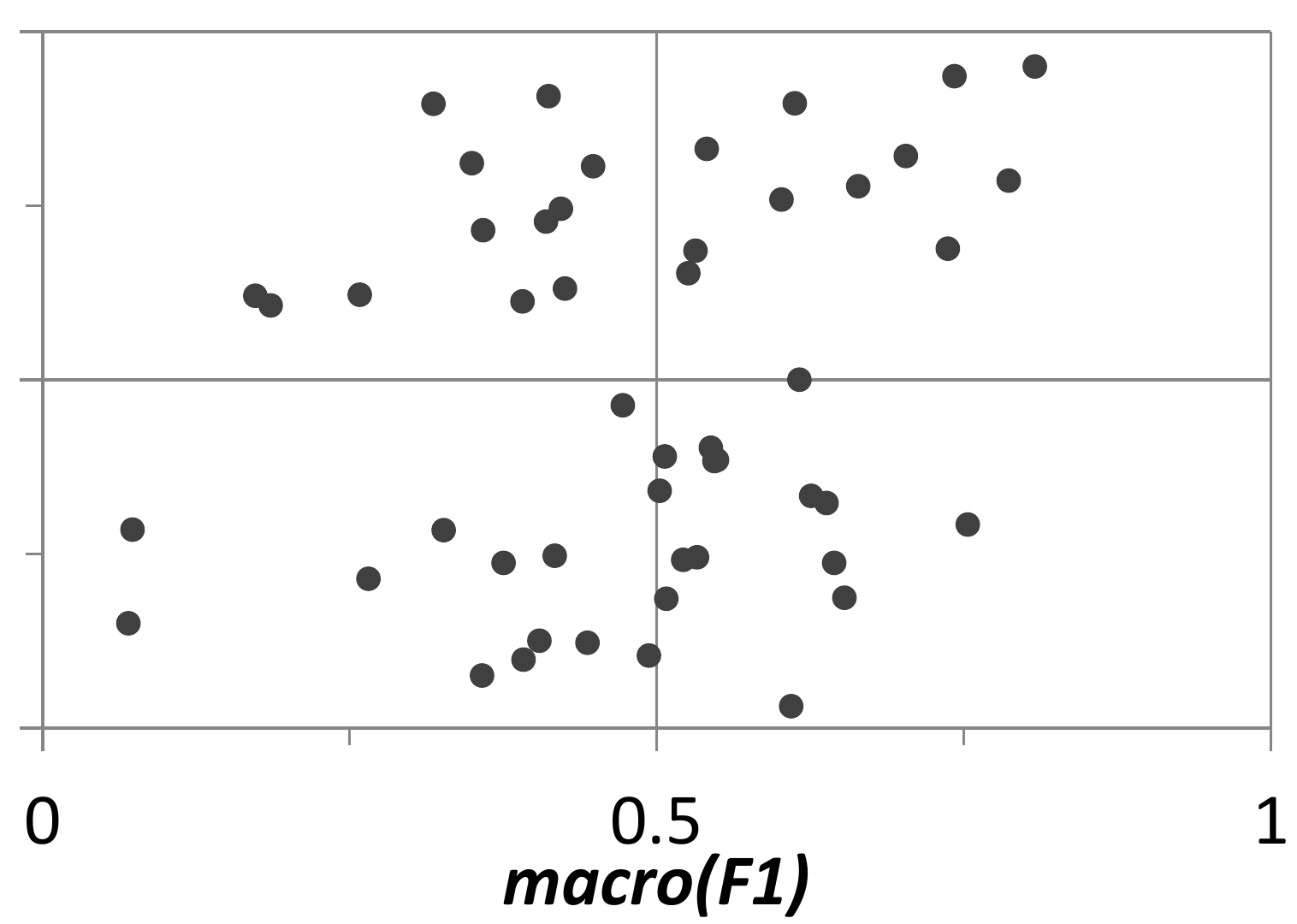} & \includegraphics[trim=130 265 60 310, clip=true, width=143pt]{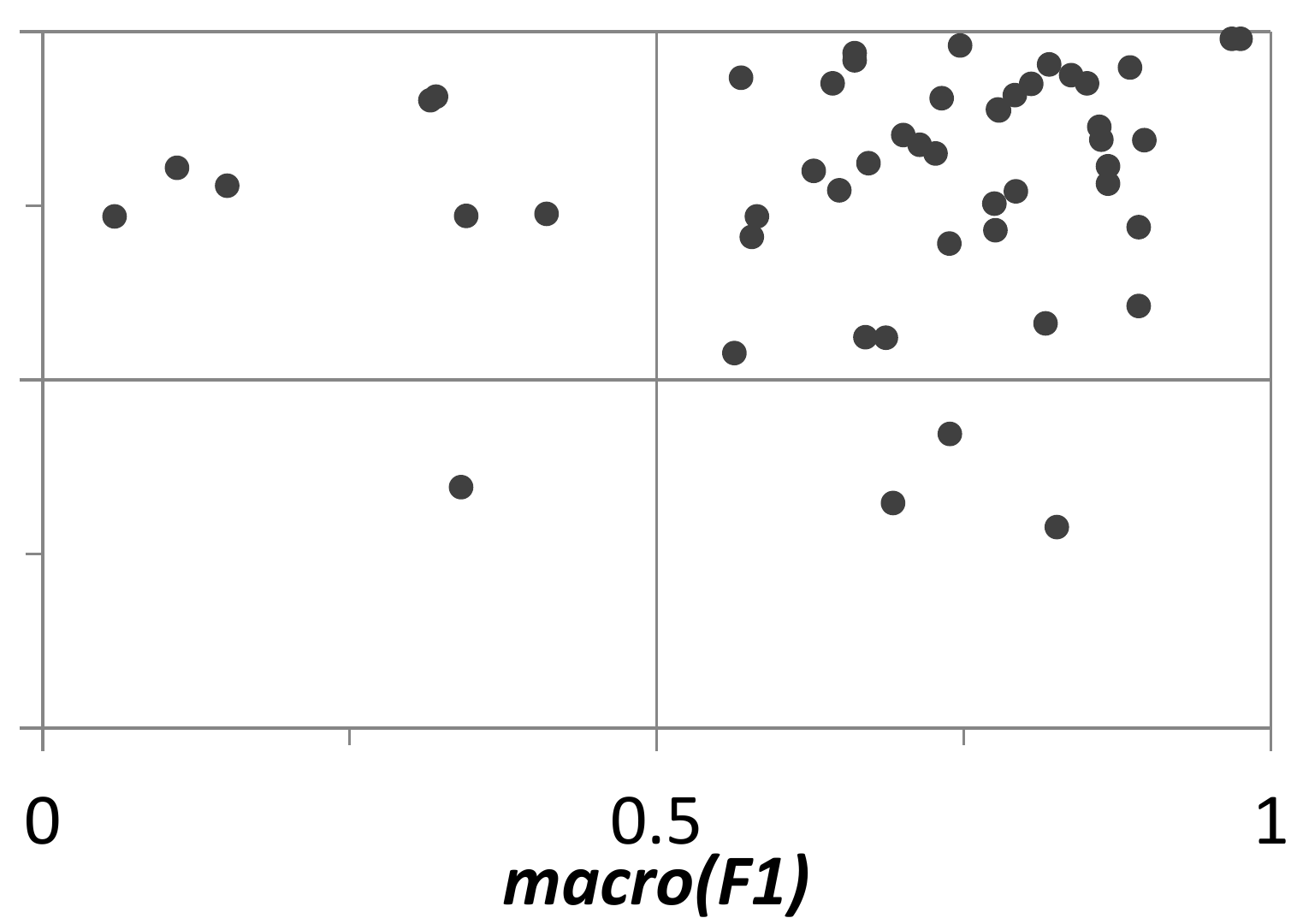} 
\\
\\
\end{array}$
\end{center}
\vspace*{-15px}
\caption{Performance comparison between different algorithm configurations. In row (a) \textSimilarityScoringFunction~-- cosine similarity, (b) \textScoringFunction~-- scoring function, (c) \textSmoothedScoringFunction~-- smoothed scoring function, (d) \textBernoulliLaplaceProb~-- Bernoulli \textNaiveBayes model with Laplace smoothing, and (e) \textMultinomialJelinekProb~-- Multinomial \textNaiveBayes model with Jelinek-Mercer smoothing; in the left-most column the configurations with no noise entity, in the middle column, the union noise profile (\textUnionAllNoiseProfile), and in the right column the intersection noise profile configurations (\textIntersectionNoiseProfile).}
\label{fig:evaluation:noise_vs_smoothing}
\end{figure*}

A commonly used measure to evaluate the performance of binary categorization methods is the \textFMeasure measure. It is defined as the harmonic mean of precision and recall and mitigates the influence of large precision or recall values. For our multi-class classification problem, we apply the micro- and macro-averaged \textFMeasure (\textMicroFMeasure and \textMacroFMeasure, respectively)~\cite{Yang1999}, which also range between~0 and~1. By the \textMicroFMeasure all documents evenly weighted, thus giving large classes a higher chance to dominate the overall score. On the other hand, the values of \textMacroFMeasure tends to be biased towards the performance of the classifier with respect to rare classes, because in this case the classes are weighted uniformly.

\begin{table}[ht]
\centering
\begin{bigtabular}{|c|c|c|c|}
\hline& & & \\[-2.2ex]
           & {\bf no noise} & {\bf union-noise} &  {\bf intersection-noise} \\
\hline& & & \\[-2.2ex]
{\boldmath$\similarityScoringFunction$\unboldmath} &      0.233 &      0.438 &      0.601 \\
\hline& & & \\[-2.2ex]
{\boldmath$\scoringFunction$\unboldmath} &      0.210 &      0.704 &      0.701 \\

{\boldmath$\smoothedScoringFunction$\unboldmath} &      0.322 &      0.569 & {\bf 0.734} \\
\hline& & & \\[-2.2ex]
{\boldmath$\bernoulliLaplaceProb$\unboldmath} &      0.251 &      0.634 &      0.642 \\

{\boldmath$\multinomialJelinekProb$\unboldmath} &      0.257 &      0.486 & {\bf 0.730} \\
\hline
\end{bigtabular}  

\caption{Performance comparison between different mapping functions for different noise entity profiles ($\overline{\fMeasure}$ scores)}
\label{tab:evaluation:noise_vs_smoothing}
\end{table}

Table~\ref{tab:evaluation:noise_vs_smoothing} compares the performance of different mapping functions and configurations for the noise entity. Each row shows the performance of one of the membership scoring functions described in Sections~\ref{sec:algorithm:vector_space_model} ($\similarityScoringFunction$, $\scoringFunction$, $\smoothedScoringFunction$) and~\ref{sec:algorithm:probabilistic_model} ($\bernoulliLaplaceProb$, $\multinomialJelinekProb$). The scoring function $\similarityScoringFunction$ serves as a baseline and refers to a simple application of the cosine similarity on the weighted features defined in Section~\ref{sec:algorithm:vector_space_model}. Each column shows the influence of different configurations for the noise entity: a union-noise entity ($\noiseProfileUnionAll$), an intersection-noise entity ($\noiseProfileIntersection$) (see Section~\ref{sec:algorithm:noise_profiles}), and a ``no noise entity'' configuration. The ``no noise entity'' column stands for a configuration without any noise entity (i.e., $E' = E$). The values in the cells represent the micro and macro \textFMeasure scores averaged over all 50 name tasks  from the previously introduced sample $\mathcal{X}$ of the evaluation dataset. More specifically:
\[
\overline{\fMeasure}=\frac{1}{|\mathcal{X}|}\sum\limits_{x \in \mathcal{X}}\frac{\microFMeasure+\macroFMeasure}{2}
\]

The influence of the noise entity profiles on the performance is notable. The configurations with the union-noise entity and the intersection-noise entity considerably outperform the ``no noise" configurations. This finding highlights the importance of mechanisms that can deal with the presence of search results that are not related to any of the source entities. Especially in our approach the noise entities are crucial for the final grouping of search results, since the bootstrapped source is of limited scope and the algorithms have to handle the open-world assumption (i.e., with results about individuals that do not occur in the underlying entity source, Wikipedia).

However, the configurations with the union-noise entity performed worse than the model represented by the mapping function \textScoringFunction (i.e., the model that quantifies the dot-product-based similarity between a result document and an entity profile). We hypothesize that this is due to the fact that the larger union-noise entity introduces too much noise for the smoothed scoring function (\textSmoothedScoringFunction, which modifies the original vectors to capture the implicit similarity between them) and also leads to a degraded performance of the Multinomial \textNaiveBayes model (\textMultinomialJelinekProb, which relies on multiple occurrences of features) since in the union-noise entity every feature occurs only once. This effect is lower for the more carefully constructed and generally smaller intersection-noise entity.

A more detailed analysis of the performance of different mapping functions and configurations for the noise entity profile is illustrated in Figure~\ref{fig:evaluation:noise_vs_smoothing}. Again, each row shows the performance of a membership scoring functions described in Sections~\ref{sec:algorithm:vector_space_model} ($\similarityScoringFunction$, $\scoringFunction$, $\smoothedScoringFunction$) and~\ref{sec:algorithm:probabilistic_model} ($\bernoulliLaplaceProb$, $\multinomialJelinekProb$). The columns show the influence of the different noise entity profiles. Moreover the charts in each cell illustrate the \textMicroFMeasure (y-axis) and \textMacroFMeasure (x-axis) of a configuration given all 50 names from the previously introduced evaluation dataset.

As one can see, the influence of the \textbf{noise entity profiles} on the performance is notable. Both configurations with noise profile (\textUnionAllNoiseProfile and \textIntersectionNoiseProfile) considerably outperform the configurations without noise profiles. Specifically the differences in the \textMicroFMeasure values are remarkable. This is due to the fact that the micro averaged \textFMeasure favors larger classes, which, given the limited scope of our bootstrapping source (i.e., \Wikipedia), are often the noise profiles. Indeed \Wikipedia contains only information about notable people; hence, there are many result documents about people having the same names as the \Wikipedia entities, which have to be assigned to the noise profiles.

The configurations with the intersection noise profile entities (\textIntersectionNoiseProfile) seem to outperform the union noise profile (\textUnionAllNoiseProfile) in all cases but for \textScoringFunction and \textBernoulliLaplaceProb where both noise profiles perform equally good.

Next, the influence of the proposed feature \textbf{smoothing solutions} on the results is compared. To this end, the basic vector-space similarity model \textScoringFunction is compared to its more advanced version \textSmoothedScoringFunction (which captures implicit similarities between documents and entity profiles, see Section~\ref{sec:algorithm:vector_space_model}). First of all, both models outperform the simple cosine similarity version (\textSimilarityScoringFunction) when a noise profile is available. This means that both versions are better suited to bootstrapping sources of limited scope. Furthermore,  \textSmoothedScoringFunction outperforms \textScoringFunction for intersection noise profile, whereas \textScoringFunction seems to perform slightly better with the union noise profile (i.e., \textMicroFMeasure). This is because the combination of the union noise profile with the implicit similarity function \textSmoothedScoringFunction introduces too much noise.

Next, the performance of the Bernoulli \textNaiveBayes model with Laplace smoothing (\textBernoulliLaplaceProb) is compared to the Multinomial \textNaiveBayes model with Jelinek-Mercer smoothing (\textMultinomialJelinekProb) (see Section~\ref{sec:algorithm:probabilistic_model}). Again, both approaches are better than \textSimilarityScoringFunction in case of a provided noise profile entity and are on a par for ``no noise entity''. One can also see, that the \textBernoulliLaplaceProb-estimator works better than \textMultinomialJelinekProb in combination with \textUnionAllNoiseProfile. However, \textMultinomialJelinekProb clearly outperforms \textBernoulliLaplaceProb in combination with \textIntersectionNoiseProfile, thus achieving the best overall performance among the probabilistic models. We hypothesize that the advantage of the Multinomial model (i.e., taking multiple occurrences of features into account) is diminished by the larger \textUnionAllNoiseProfile, in which all the features occur only once. This impact is much smaller in the case of \textIntersectionNoiseProfile.

The best results are thus achieved by \textSmoothedScoringFunction and \textMultinomialJelinekProb-estimator in combination with the \textIntersectionNoiseProfile. Furthermore, the baseline algorithm \textSimilarityScoringFunction performs best with \textIntersectionNoiseProfile too.


\section{Conclusion}
\label{sec:conclusion}
The focus of this work has been on the design of efficient and high-quality approaches to the problem of clustering search results to ambiguous person-name queries. The proposed methods build on the idea that Web~2.0 profiles, can be bootstrapped to cast the above problem into a classification problem, where results are mapped to the most similar profile. The proposed method does not influence the grouping of entities not represented in the knowledge base, which might be clustered subsequently. With a steady increase of information about people and other entities in social networks and other Web~2.0 sources, we are confident that methods like the ones presented in this paper will gain importance for reliably disambiguating search results.

The suite of presented and evaluated methods covers a wide range of advanced vector-space and probabilistic models. Efficient similarity and likelihood estimations that do not compromise the classification quality have been a primary goal of this work. Furthermore, dealing with noisy and biased data from the Web documents as well as the knowledge base was essential for the introduced approach. Specifically, handling the incompleteness of the entity source (open world assumption) was as well in the focus of this work.

The provided experiments were based on a hand-labeled dataset over more than $85k$ alignment candidates of around 5,000 Web pages on ambiguous person names that we have made publicly available. Although all methods deliver satisfactory results, in light of the experimental outcome, we would favor the smoothed vector space model implementing \textSmoothedScoringFunction. For a definitive answer all methods would have to be evaluated on multiple datasets.

Note that all presented methods can be analogously applied to a broader online scenario, where for a given ambiguous query, Web results and appropriate entity profiles would be retrieved in parallel and search results mapped to the most similar profile. Although there would be a moderate increase in latency due to online indexing of unigrams in documents and profiles, we expect the result quality to outweigh the runtime latencies.

As part of our future work, we are aiming to aggregate the results for different Web~2.0 sources (e.g., LinkedIn, Facebook, etc.) to improve the grouping of search results to ambiguous queries. To this end, more complex features (e.g., multigrams, structured attribute-value pairs, etc.) could boost our methods further. An incremental classification of search results is an additional point on our agenda. One could first map only those result documents to profiles for which the similarity exceeds some threshold and then given the gained information, hopefully a better classification of the remaining documents would be achieved. Obviously, there is high value in the information accessible through Web~2.0 APIs and future Web search should aim at better exploiting this information.

\balance

\bibliographystyle{abbrv}
\bibliography{references}

\end{document}